\documentclass[onecolumn,10pt]{emulateapj}
\usepackage{times}

\newcommand{\D}[2]{\frac{\partial #2}{\partial #1}}

\newcommand\bb[1] {\mbox{\boldmath{$#1$}}}
\newcommand\del{\bb{\nabla}} \newcommand\bcdot{\bb{\cdot}}
 \newcommand\beq{\begin{equation}}
\newcommand\eeq{\end{equation}} \newcommand{\Alfven}{Alfv\'{e}n }
\newcommand{\Alfvennospace}{Alfv\'{e}n}

\begin{document}
%\submitted{Draft, \today}
\shorttitle{\textsc{MAGNETIZED ROTATING PLASMAS WITH SUPERTHERMAL FIELDS}}
\shortauthors{\textsc{PESSAH \& PSALTIS}}

\title{\textsc{THE STABILITY OF MAGNETIZED ROTATING PLASMAS WITH
    SUPERTHERMAL FIELDS}} 

%\received{} \revised{}    \accepted{}

\author{\textsc{Martin E. Pessah}\altaffilmark{1,2} and
\textsc{Dimitrios Psaltis}\altaffilmark{2,1}}
\altaffiltext{1}{Astronomy Department, University of Arizona, 933
N. Cherry Ave., Tucson, AZ, 85721; mpessah@as.arizona.edu }
\altaffiltext{2}{Physics Department, University of Arizona, 1118 E.
4th St., Tucson, AZ, 85721; dpsaltis@physics.arizona.edu}

\begin{abstract}
During the last decade it has become evident that the
magnetorotational instability is at the heart of  the enhanced angular
momentum transport in weakly magnetized accretion disks around neutron
stars and black holes.  In this paper, we investigate the local linear
stability of differentially rotating, magnetized flows and the
evolution of the magnetorotational instability beyond the weak-field
limit. We show that, when superthermal  toroidal fields are
considered,  the effects of both compressibility and magnetic tension
forces, which are related to  the curvature of toroidal field lines,
should be taken fully into account.  We demonstrate that 
the presence of a strong toroidal component
in the magnetic field plays a non-trivial role. When strong fields are
considered, the strength of the toroidal  magnetic field not only
modifies the growth rates of the unstable modes but also determines
which modes are subject to instabilities. We find that, for
rotating configurations with Keplerian laws,
the magnetorotational instability is stabilized at low wavenumbers for
toroidal \Alfven speeds exceeding the geometric mean of the sound
speed and the rotational speed. For a broad range of magnetic field
strengths, we also find that two additional distinct instabilities are
present; they both appear as the result of coupling between the modes
that become the \Alfven and the slow modes in the limit of no
rotation. We discuss the significance of our  findings for the
stability of cold, magnetically dominated, rotating fluids and argue
that, for these systems, the curvature of toroidal field lines cannot
be neglected even when short wavelength perturbations are considered.
We also comment on the implications of our results for the validity of
shearing box simulations in which superthermal toroidal fields are
generated.
\end{abstract}

\keywords{accretion, accretion disks --- MHD --- instabilities ---
plasmas}

\section{\textsc{INTRODUCTION}}
Linear mode analyses provide a useful  tool in gaining important
insight into the relevant physical  processes determining the
stability of magnetized accretion flows. Studies of local linear modes
of accretion disks threaded by weak magnetic fields have pointed out
important clues on viable mechanisms for angular momentum transport
and the subsequent accretion of matter onto the central objects
\citep{BH91, BH98, BH02, SM99, Balbus03}. They have also provided
simplified physical models and analogies over which more complex
physics can, in principle, be added \citep{BH92, BH98, Quataert02}.
These treatments were carried out in the magnetohydrodynamic (MHD)
limit  (but see Quataert, Dorland \& Hammett 2002 who studied the
kinetic limit) and invoked a number of approximations appropriate to
the study of the evolution of short-wavelength perturbations, when
weak fields are considered. In this context, the strength of the
magnetic field, $B$, is inferred by comparing the thermal pressure,
$P$,  to the magnetic pressure and is characterized by a plasma
parameter, $\beta \equiv 8\pi P/ B^2 > 1$.

It is not hard to find situations of astrophysical interest, however,
in which the condition of weak magnetic fields is not satisfied.  A
common example of such a situation is the innermost  region of an
accretion disk around a magnetic neutron star.  It is widely accepted
that X-ray pulsars are powered by accretion of matter onto the polar
caps of magnetic neutron stars.  For this to occur, matter in the
nearly Keplerian accretion disk has to be funneled along the field
lines. This suggests that, at some radius, centrifugal forces and
thermal pressure have to be overcome by magnetic stresses, leading
naturally to regions where $\beta \lesssim 1$.

In the context of accretion disks, the presence of superthermal fields
in rarefied coronae also seems hard to avoid, if coronal heating is a
direct consequence of the internal dynamics of the disk itself rather
than being produced by external irradiation from the central
object. Three-dimensional MHD simulations by \citet{MS00} showed  that
magnetic turbulence can effectively couple with buoyancy to transport
the  magnetic energy produced by the magnetorotational instability
(MRI) in weakly magnetized disks to create a strongly magnetized
corona within a few scale heights from the disk plane.   On long time
scales, the average vertical disk structure consists of a weakly
magnetized ($\beta \simeq 50$) turbulent core below two scale heights
and a strongly magnetized ($\beta \lesssim 0.1$) non-turbulent corona
above it. The late stages of evolution in these models show that the
disks themselves become magnetically dominated.  \citet{MHM00} also
found that the average plasma $\beta$ in  disk coronae is $\simeq
0.1-1$ and the volume filling factor for regions with $\beta \lesssim
0.3$ is up to $0.1$.  Even in the absence of an initial toroidal
component, simulations carried out by \citet{KMS02} showed that
low-$\beta$ regions develop near the equator of the disk  because of a
strong toroidal component of the magnetic field generated by shear.
On more theoretical grounds, strong toroidal magnetic fields produced
by strong shear in the boundary layer region  have been suggested as
responsible for the observed bipolar outflows in young stellar objects
\citep{Pringle89}.  More recently, \citet{PBB03} found self-consistent
solutions for thin magnetically-supported accretion disks and pointed
out the necessity of assessing the stability properties of such
configurations.

Another case in which magnetic fields seem to play an important
dynamical role in rotating fluid configurations is that of
magnetically supported molecular clouds.  Observations of both large
Zeeman line-splitting and of broad molecular  lines support
the presence of superthermal fields 
(see Myers \& Goodman 1988 for further references
and Bourke \& Goodman 2003 for a review on the current understanding
on the role of magnetic fields in molecular clouds).  Values of the
plasma $\beta$ of the order of $0.1-0.01$ have also been used in
numerical studies of the structural properties of giant molecular
clouds \citep{OSG01}.

As a last example of astrophysical interest,  we mention
magnetocentrifugally driven winds,  such as those observed  in
protostars. These outflows seem to play an important role in the
evolution of young  stellar objects and in the dynamics of the parent
clouds  by providing a source of turbulent energy.  Magnetocentrifugal
jets typically involve internal \Alfven speeds comparable to the flow
speeds.   These structures are supported mainly by magnetic pressure
due to strong toroidal fields. The ratio of magnetic pressure in the
jet to the gas pressure of the ambient medium can be of the order of
$10^6$ (for an extensive study of MHD driven instabilities in these
systems see Kim \& Ostriker 2000 and \S \ref{subsec:Comparison to
previous analytical studies}).

In this paper, we investigate the local linear stability of
differentially rotating flows without imposing any \emph{a priori}
restrictions on the strength of the magnetic field. We do, however,
restrict our attention to rotationally supported flows 
(we loosely use this term to refer to flows with
internal \Alfven speeds smaller than the rotational speed). Our intent
is to demonstrate that the effects of the finite curvature of the
toroidal field lines on the stability of small-wavelength vertical
perturbations (i.e., on the most unstable modes present in the
weak-field MRI) cannot be neglected when  superthermal toroidal fields
are present.  In order to achieve this task, we relax the Boussinesq
approximation (see also Papaloizou \& Szuszkiewicz 1992 and Blaes \&
Balbus 1994), which is valid only when the toroidal component of the
field is subthermal \citep{BH91}. We thus consider the MHD fluid to be
fully compressible. Moreover, even though we perform a local analysis,
we do consider curvature terms when evaluating magnetic forces, for
they become important in the strong-field regime (see also Knobloch
1992 and Kim \& Ostriker 2000).

In most early studies addressing the MRI,  it was found that the only
role played by a toroidal component in the magnetic field is to quench
the growth rates of the modes that are already unstable when only weak
vertical fields are considered (Balbus \& Hawley 1991; Blaes \& Balbus
1994; see Quataert, Dorland,  \& Hammett 2002 for the kinetic limit;
see also Kim \& Ostriker 2000  for the cold MHD limit). Here, we show
that, when strong fields are considered and the approximations usually
invoked in the study of the weak-field MRI  are relaxed, the presence
of a toroidal component of the magnetic field plays a crucial role not
only in the growth rates of the unstable modes but also in determining
which modes are subject to instabilities\footnote{We will comment
later in more detail on the paper by \citet{CP95} who outlined the
effects of a dynamically important toroidal field (in the case of an
incompressible MHD flow) and address the similitudes and differences
with our findings.}. As expected, the presence of a toroidal component
breaks the symmetry of the problem, also giving rise to traveling
modes. Moreover, for a broad range of magnetic field strengths, we
find that two different instabilities are present. They both appear as
the result of coupling between the modes that become the \Alfven and
the slow mode in the limit of no rotation.

The paper is organized as follows. In \S  \ref{sec:mhd equations and
dispersion relation},  we describe the physical setup to be studied,
present the dispersion relation to be solved, and discuss the
importance of curvature terms in the limit of superthermal fields.  In
\S \ref{sec:numerical solutions}, we solve numerically the dispersion
relation in some interesting regimes.   In \S \ref{sec:onset of
instabilities},  we study the onset of instabilities as a function of
magnetic field strengths and present some useful approximate criteria
that enable us to study analytically some aspects of the full
problem. In \S \ref{sec:discussion}, we compare our results to
previous investigations and discuss some of the implications of this
study. Finally, in  \S \ref{sec:summary and conclusions}, we present a
brief summary and our conclusions.

\section{\textsc{MHD EQUATIONS FOR PERTURBATIONS AND THE DISPERSION RELATION}}
\label{sec:mhd equations and dispersion relation}

We start with the set of equations that govern the behavior of a
polytropic fluid in the MHD approximation,
\begin{equation}
\label{eq:continuity}
\D{t}{\rho} + \del\bcdot(\rho\bb{v}) = 0 ~,
\end{equation}
\begin{eqnarray}
\label{eq:euler}
\rho\D{t}{\bb{v}} + \left(\rho\bb{v}\bcdot\del\right)\bb{v} =  -
\rho\del\Phi - \del\left(P + \frac{\bb{B}^2}{8\pi}\right)  +
\left(\frac{\bb{B}}{4\pi}\bcdot\del\right)\bb{B} ~,
\end{eqnarray}
\begin{equation}
\label{eq:induction}
\D{t}{\bb{B}} + \left( \del \bcdot \bb{v} \right) \bb{B} - \left(
\bb{B} \bcdot \del \right) \bb{v} + \left( \bb{v} \bcdot \del \right)
\bb{B} = 0 ~,
\end{equation}
and
\begin{equation}
\label{eq:energy}
P = P_0 \left(\frac{\rho}{\rho_0} \right)^\Gamma ~.
\end{equation}
In these equations, $\rho$ is the mass density, $\bb{v}$ the velocity,
$P$ the gas pressure, and $\Gamma$ the polytropic index; $\bb{B}$ is
the magnetic field and $\Phi$ the gravitational potential.   For
convenience we adopt a cylindrical set of coordinates $(r,\phi,z)$
with origin in the central object  (i.e., the neutron star or black
hole).  We assume a steady axisymmetric background flow
characterized by a velocity field of the form  $\bb{v} = v_{\phi}(r,
z) \hat \phi$ and threaded by a background magnetic field. For
consistency, our analysis is restricted to background fields of the
form $\bb{B} = B_{\phi} \hat \phi + B_z \hat z$ since the effect of
including a radial component in the field is to generate a linear
growth in time of the toroidal component \citep{BH91}.  Under these
circumstances, all the background quantities depend on the radial and
vertical coordinates only.   In the present treatment, we neglect the
self gravity of the fluid.  In fact, \citet{PBB03} showed that
magnetically dominated accretion disks have lower surface and volume
densities for a fixed accretion rate. This suggests that these systems
are  lighter than standard disks and thus are not subject to
self-gravity instabilities.

\subsection{Equations for the Perturbations}
\label{sec:equations for the perturbations}

In order to perform the local linear
mode analysis, we perturb the set of equations
(\ref{eq:continuity})-(\ref{eq:energy}) by substituting every physical
variable $f$ by $f + \delta f$ and retain only linear orders in
$\delta f$.   In the following, we focus our analysis on the study of
axisymmetric perturbations in an axisymmetric background.

We first consider, in some detail, the  radial component  of the
momentum equation (\ref{eq:euler}), which becomes
\begin{eqnarray}
\label{eq:euler_r_linear_step1}
\D{t}{\delta v_r} &-& 2\Omega \delta v_\phi + \frac{1}{\rho}  \left[
\D{r}{\delta P} + \frac{1}{4\pi}  \left( \D{r}{B_\phi} \delta B_\phi +
B_\phi  \D{r}{\delta B_\phi} + \D{r}{B_z} \delta B_z + B_z
\D{r}{\delta B_z} \right)\right] -\D{r}{P} \frac{\delta \rho}{\rho^2}
\nonumber \\ &-&\frac{1}{4\pi\rho} \left( B_\phi  \D{r}{B_\phi}
+ B_z  \D{r}{B_z} \right) \frac{\delta \rho}{\rho} -
\frac{1}{4\pi\rho} \left[ B_z \D{z}{\delta B_r}   - 2 \frac{B_\phi}{r}
\delta B_\phi \right] - \frac{1}{4\pi\rho}  \frac{B_\phi^2}{r}
\frac{\delta \rho}{\rho} =0 ~.%\nonumber \\
\end{eqnarray}
The coefficients in this linear equation for the perturbed
variables depend in general on $r$ and $z$,  [e.g., the angular
velocity $\Omega$ is in general $\Omega(r,z)$].  Therefore, at this
point, the decomposition of perturbed quantities in Fourier modes --
e.g., symbolically $\delta f = \sum \delta f_k  e^{i(k_r r+k_z z -
\omega t)}$ -- would not result in any particular simplification of
the problem.  This can be seen by taking the Fourier transform of
equation (\ref{eq:euler_r_linear_step1}) which results in a sum of
convolutions of the Fourier transforms of background and perturbed
quantities.  Further progress can be made if we restrict the
wavelengths of the perturbations for which our stability analysis is
valid. To this end, we choose a fiducial point, $\bb{r}_0 = (r_0,
\phi_0, z_0)$, around which we perform the local stability analysis.
The choice of the particular value of $\phi_0$ is, of course,
irrelevant in the axisymmetric case under study.

We expand all the background quantities in equation
(\ref{eq:euler_r_linear_step1}) in Taylor series around $\bb{r}_0$ and
retain only the zeroth order in terms of the local coordinates
$\xi_r=r-r_0$ and $\xi_z=z-z_0$ to obtain
\begin{eqnarray}
\label{eq:euler_r_linear_step2}
\D{t}{\delta v_r}  &-& 2\Omega_0 \delta v_\phi  + \frac{1}{\rho_0}
\left[ \D{\xi_r}{\delta P}  + \frac{1}{4\pi} \left(
  \left.\D{r}{B_\phi}\right|_0 \delta B_\phi  + B^0_\phi
  \D{\xi_r}{\delta B_\phi} +  \left.\D{r}{B_z}\right|_0 \delta B_z  +
  B^0_z  \D{\xi_r}{\delta B_z} \right)\right] -
\left.\D{r}{P}\right|_0 \frac{\delta \rho}{\rho_0^2} \nonumber \\
&-&\frac{1}{4\pi\rho_0} \left( B^0_\phi  \left. \D{r}{B_\phi} \right|_0 +
B^0_z  \left.\D{r}{B_z} \right|_0\right) \frac{\delta \rho}{\rho_0} -
\frac{1}{4\pi\rho_0} \left( B^0_z \D{\xi_z}{\delta B_r}   - 2
\frac{B^0_\phi}{r_0} \delta B_\phi \right) - \frac{1}{4\pi\rho_0}
\frac{(B^0_\phi)^2}{r_0} \frac{\delta \rho}{\rho_0} =0 ~.%\nonumber \\
\end{eqnarray}
Here, $\Omega_0$, $\rho_0$, $B_\phi^0$, and $B_z^0$  stand for the
angular velocity,  background density, and magnetic field components
at the fiducial point $\bb{r}_0$ and the subscript ``$0$'' in the
derivatives with respect to the radial coordinate $r$ indicates that they
are evaluated at $\bb{r}_0$.  Equation (\ref{eq:euler_r_linear_step2})
is a linear partial-differential  equation in the local variables for
the perturbed quantities  but with constant coefficients.  This is a
good approximation as long as the departures $(\xi_r, \xi_z)$ are
small compared to the length scales over which there are significant
variations in the background quantities, i.e,  $\xi_r\ll L_r$ and
$\xi_z\ll L_z$, where $L_r$ and $L_z$ are the characteristic length
scales in the radial and vertical directions, respectively.

It is only now that it is useful to expand the perturbed quantities
in equation (\ref{eq:euler_r_linear_step2}) in Fourier modes.  We can thus
write each one of the perturbed quantities as
\begin{equation}
\delta f = \delta   f(k_r, k_z, \omega)  \, e^{i(k_r \xi_r+k_z \xi_z -
\omega t)}  ~,
\end{equation}
and write the radial momentum equation for each mode as
\begin{eqnarray}
\label{eq:euler_r_linear_step3}
- i \omega \delta   v_r &-& 2\Omega_0 \delta   v_\phi  - ik_z
\frac{B^0_z \delta   B_r}{4\pi\rho_0} + \left( ik_r + \frac{2}{r_0} +
\left.\D{r}{\ln B_\phi}\right|_0 \right) \frac{B_\phi^0 \delta
  B_\phi}{4\pi\rho_0} + \left( ik_r + \left.\D{r}{\ln B_z}\right|_0
\right)  \frac{B_z^0 \delta   B_z}{4\pi\rho_0} \nonumber \\ &+& \left
\{ \left(ik_r -  \left.\D{r}{\ln \rho}\right|_0 \right)  (c^0_{\rm
  s})^2  -\left[ \left( \frac{1}{r_0}
  +\left.\D{r}{\ln B_\phi}\right|_0 \right) (v^0_{{\rm A} \phi})^2 +
  \left.\D{r}{\ln B_z}\right|_0  (v^0_{{\rm A} z})^2 \right] \right \}
\frac{\delta \rho}{\rho_0}=0 ~.%\nonumber \\
\end{eqnarray}
Here, we have introduced the quantities $c^0_{\rm s}, v^0_{{\rm A}
\phi}$ and  $v^0_{{\rm A} z}$ that stand for the local sound speed and
local \Alfven speeds associated with the toroidal and vertical
components of the local magnetic field and are defined by
\begin{equation}
\label{eq:speeds}
c^0_{\rm s} \equiv \sqrt{\Gamma \frac{P_0}{\rho_0}} ~, \qquad
v^0_{{\rm A}\phi} \equiv  \frac{B^0_\phi}{\sqrt{4\pi \rho_0}} ~,
\qquad \textrm{and} \qquad v^0_{{\rm A} z} \equiv
\frac{B^0_z}{\sqrt{4\pi \rho_0}}  ~.
\end{equation}
Note that, for brevity, we have omitted the dependences $(k_r, k_z,
\omega)$ in the Fourier amplitudes. For consistency, the validity of
the analysis  is now restricted to modes with wavenumbers satisfying
$k_r L_r \gg 1$ and $k_z L_z \gg 1$.  Without loss of generality, we
assume that the fiducial point is inside the disk so we can write
the local conditions on the wavenumbers as $k_r r_0 \gg 1$ and $k_z
z_0 \gg 1$ . Moreover, for fiducial points such that $r_0 \ge z_0$ the
latter condition  also implies $k_z r_0 \gg 1$.

At this point, it is also convenient to define a new set of
independent variables  $(\delta v_{{\rm A} r}, \delta v_{{\rm A}
\phi},\delta v_{{\rm A} z} )$ defined in terms of $(\delta B_r, \delta
B_\phi, \delta B_z)$ in such a way that $\delta \bb{v}_{A} \equiv \delta
\bb{B}/\sqrt{4\pi \rho_0}$.  In this case, equation
(\ref{eq:euler_r_linear_step3}) reads
\begin{eqnarray}
\label{eq:euler_r_linear_step4}
- i \omega \delta v_r &-& 2\Omega_0 \delta v_\phi  - ik_z v^0_{{\rm A}
  z} \delta v_{{\rm A} r} + \left( ik_r + \frac{2}{r_0} +
\left.\D{r}{\ln B_\phi}\right|_0 \right) v^0_{{\rm A} \phi} \delta
v_{{\rm A} \phi} + \left( ik_r + \left.\D{r}{\ln B_z}\right|_0 \right)
v^0_{{\rm A} z} \delta v_{{\rm A} z} \nonumber \\ &+& \left \{
\left(ik_r -  \left.\D{r}{\ln \rho}\right|_0 \right) (c^0_{\rm s})^2 -
\left[ \left( \frac{1}{r_0} + \left.\D{r}{\ln B_\phi}\right|_0 \right)
  (v^0_{{\rm A} \phi})^2  +  \left.\D{r}{\ln B_z}\right|_0 (v^0_{{\rm
      A} z})^2\right] \right \}\frac{\delta \rho}{\rho_0}=0 ~.%\nonumber \\
\end{eqnarray}

As a last step, it is useful to work with dimensionless quantities.
To this end, we define dimensionless variables by scaling  all the
frequencies with the local rotational frequency $\Omega_0$ and all
speeds with the local circular velocity $\Omega_0 r_0$.  It is also
convenient to define  dimensionless wavenumbers by multiplying the
physical wavenumber by the radial coordinate $r_0$.  In summary, we
define
\begin{eqnarray}
\tilde \omega  = \omega/\Omega_0 ~,  \qquad  &\tilde k_r = k_r r_0 ~,&
\qquad  \tilde k_z = k_z r_0 ~, \\ \tilde c^0_{\rm s} = c^0_{\rm
s}/\Omega_0 r_0 ~,  \qquad  &\tilde v^0_{{\rm A} \phi} =  v^0_{{\rm A}
\phi}/\Omega_0 r_0 ~, &\qquad  \tilde v^0_{{\rm A} z} =  v^0_{{\rm A}
z}/\Omega_0 r_0 ~, \\ \delta \tilde  \rho = \delta \rho/\rho_0 ~,
\qquad  &\delta \tilde{\bb{v}} = \delta \bb{v}/\Omega_0 r_0 ~,& \qquad
\delta \tilde{\bb{v}}_{\rm A} = \delta \bb{v}_{\rm A}/\Omega_0 r_0 ~.
\end{eqnarray}
For completeness,  we introduce here the epicyclic frequency $\kappa$
and its local dimensionless counterpart
\begin{equation}
\label{eq:epicyclic}
\kappa = 2\Omega \left[1 + \frac{1}{2}\frac{d\ln \Omega}{d\ln
    r}\right]^{1/2} \qquad \textrm{and} \qquad \tilde{\kappa}_0 =
    \frac{\kappa_0}{\Omega_0}  = 2 \left[1 + \frac{1}{2}
    \left. \frac{d\ln \Omega}{d\ln r} \right|_0  \right]^{1/2} ~.
\end{equation}
This quantity appears naturally in stability analyses of
differentially rotating configurations and it is the frequency at
which all the flow variables oscillate around their background values
in the absence of magnetic fields. For a rotational profile given by a
power law, the epicyclic frequency is proportional to the angular
frequency at all radii.

Finally, the dimensionless version of equation
(\ref{eq:euler_r_linear_step4}) reads,
\begin{eqnarray}
\label{eq:euler_r_linear_step_5}
- i \tilde \omega \delta \tilde v_r &-& 2\delta \tilde  v_\phi  - i
\tilde k_z \tilde v^0_{{\rm A} z} \delta \tilde v_{{\rm A} r} + \left(
i\tilde k_r + 2 + \left.\D{\ln r}{\ln B_\phi}\right|_0 \right) \tilde
v^0_{{\rm A} \phi} \delta \tilde v_{{\rm A} \phi} + \left( i\tilde k_r
+ \left.\D{\ln r}{\ln B_z}\right|_0 \right)  \tilde v^0_{{\rm A} z}
\delta \tilde v_{{\rm A} z} \nonumber \\ &+& \left \{ \left(i\tilde
k_r -  \left.\D{\ln r}{\ln \rho}\right|_0 \right) (\tilde c^0_{\rm
  s})^2 - \left[ \left( 1 + \left.\D{\ln r}{\ln B_\phi}\right|_0
  \right) (\tilde v^0_{{\rm A}\phi})^2  +  \left.\D{\ln r}{\ln
    B_z}\right|_0 (\tilde v^0_{{\rm A} z})^2\right] \right \}\delta
\tilde \rho =0 ~.%\nonumber \\
\end{eqnarray} 

Following a similar procedure with the remaining equations in the
system (\ref{eq:continuity})-(\ref{eq:energy}),  we arrive  to the
linear set of equations needed to perform the local stability
analysis. For brevity, we now drop the hat in all the dimensionless
variables and  the superscript in $\tilde{c}^0_{\rm s},
\tilde{v}^0_{{\rm A} \phi}$ and  $\tilde{v}^0_{{\rm A} z}$. We then
write
\begin{eqnarray}
\label{eq:pert_cont_grads}
- i  \omega \delta  \rho + \left( i k_r + \epsilon_4 + \left.\D{\ln
  r}{\ln \rho}\right|_0 \right) \delta   v_r +  \left( i k_z  +
\frac{r_0}{z_0}\left.\D{\ln z}{\ln \rho}\right|_0 \right) \delta v_z
=0 ~,
\end{eqnarray}
\begin{eqnarray}
\label{eq:pert_euler_r_grads}
- i  \omega \delta  v_r &-& 2\delta   v_\phi  - i  k_z  v_{{\rm A} z}
\delta  v_{{\rm A} r} + \left( i k_r + 2\epsilon_1 + \left.\D{\ln
  r}{\ln B_\phi}\right|_0 \right)  v_{{\rm A} \phi} \delta  v_{{\rm A}
  \phi} + \left( i k_r + \left.\D{\ln r}{\ln B_z}\right|_0 \right)  v
_{{\rm A} z} \delta  v_{{\rm A} z} \nonumber \\ &+& \left \{ \left(i
k_r -  \left.\D{\ln r}{\ln \rho}\right|_0 \right)  c_{\rm s}^2 -
\left[ \left( \epsilon_2 + \left.\D{\ln r}{\ln B_\phi}\right|_0
  \right) v_{{\rm A} \phi}^2  +  \left.\D{\ln r}{\ln B_z}\right|_0
  v_{{\rm A} z}^2\right] \right \}\delta  \rho =0 ~,
\end{eqnarray}
\begin{eqnarray}
\label{eq:pert_euler_phi_grads}
- i  \omega \delta  v_\phi + \frac{ \kappa^2}{2} \delta   v_r +
\frac{r_0}{z_0}  \left.\D{\ln z}{\ln \Omega}\right|_0  \delta   v_z
&-&  \left( \epsilon_3 + \left.\D{\ln r}{\ln B_\phi}\right|_0  \right)
v_{{\rm A} \phi}  \delta  v_{{\rm A} r}  \nonumber \\ &-& i k_z
v_{{\rm A} z} \delta v_{{\rm A} \phi} - \frac{r_0}{z_0} \left.\D{\ln
  z}{\ln B_\phi}\right|_0  v_{{\rm A} \phi}  \delta  v_{{\rm A} z} =
0~, 
\end{eqnarray}
\begin{eqnarray}
\label{eq:pert_euler_z_grads}
- i  \omega \delta  v_z 
&-&\left.\D{\ln r}{\ln B_z}\right|_0  v_{{\rm A} z} 
\delta  v_{{\rm A} r} + \left( i k_z +
\frac{r_0}{z_0}\left.\D{\ln z}{\ln B_\phi}\right|_0  \right) v_{{\rm
A} \phi} \delta  v_{{\rm A} \phi} \nonumber  \\ &+& \left[ \left(i
k_z -  \frac{r_0}{z_0}\left.\D{\ln z}{\ln \rho}\right|_0 \right)
c_{\rm s}^2 - \frac{r_0}{z_0} \left(\left.\D{\ln z}{\ln B_\phi}
\right|_0  v_{{\rm A} \phi}^2  
+  \left.\D{\ln z}{\ln B_z}\right|_0 v_{{\rm A} z}^2 \right)
\right] \delta  \rho =0 ~, \nonumber \\
\end{eqnarray}
\begin{equation}
\label{eq:pert_induc_r_grads}
i \omega \delta v_{{\rm A} r} + ik_z v_{{\rm A} z} \delta v_r = 0 ~,
\end{equation}
\begin{eqnarray}
\label{eq:pert_induc_phi_grads}
- i \omega \delta v_{{\rm A} \phi}  
- \left.\frac{d \ln \Omega}{d \ln r} \right|_0  \delta v_{{\rm A} r}  
- \left.\frac{r_0}{z_0}\frac{d \ln \Omega}{d \ln z}\right|_0  \delta v_{{\rm A} z}  
&+& \left(i k_r + \left.\D{\ln r}{\ln B_\phi}\right|_0 \right)  v_{{\rm A} \phi}
\delta  v_r - i k_z  v_{{\rm A} z} \delta v_\phi  \nonumber  \\ &+& \left(i k_z +
\frac{r_0}{z_0}\left.\D{\ln z}{\ln B_\phi}\right|_0 \right)  
v_{{\rm A} \phi} \delta  v_z =0 ~,  
\end{eqnarray}
and
\begin{equation}
\label{eq:pert_induc_z_grads}
- i \omega \delta v_{{\rm A} z} + \left(ik_r + \epsilon_4   +
\left.\D{\ln r}{\ln B_z}\right|_0 \right) v_{{\rm A} z} \delta v_r = 0
~,
\end{equation}
where we have used  equation (\ref{eq:energy}) to recast the pressure
perturbations in terms of density perturbations.  The local conditions
over the wavenumbers now read $k_r \gg 1$ and $k_z \gg 1$.

The factors $\epsilon_i$, with $i=1,2,3,4$, are just convenient dummy
variables that we introduce in order to help us keep track of the
terms that account for the finite curvature  of the background  and
are usually neglected in local studies of the weak-field MRI. Their
numerical values are to be regarded as unity, unless otherwise
mentioned.   The terms proportional to $\epsilon_1$ and $\epsilon_2$
in equation (\ref{eq:pert_euler_r_grads}) and the term  proportional
to $\epsilon_3$ in equation (\ref{eq:pert_euler_phi_grads}) are due to
the effects of magnetic tension and they appear naturally when a
cylindrical coordinate system is adopted. The terms proportional to
$\epsilon_4$ in equations  (\ref{eq:pert_cont_grads}) and
(\ref{eq:pert_induc_z_grads}) are related to flux conservation in
cylindrical coordinates.   Although the three terms labeled by
$\epsilon_1$, $\epsilon_2$, and $\epsilon_3$ share the same physical
origin (i.e., magnetic tension introduced by the curvature of toroidal
field lines), it is useful to be able to distinguish among  them
because the one labeled with $\epsilon_2$ vanishes in the limit of an
incompressible flow.   Note that  equations
(\ref{eq:pert_induc_r_grads}) and (\ref{eq:pert_induc_z_grads}) ensure
a divergence-free perturbed magnetic field, i.e., 
$\del\bcdot \delta \bb{B} =0$, only when the finite curvature of 
the background is  accounted for (i.e., $\epsilon_4=1$).

Up to this point, our intention has been to keep the discussion as
general as possible in order to clearly state all the assumptions that
we have made to obtain the set of equations for the
perturbations to perform a  local linear mode analysis. For the sake
of simplicity, and to avoid the parametric study from being too
extensive, we further invoke the following assumptions. We choose the
fiducial point $\bb{r}_0$ to lie in the disk mid-plane and assume
that, locally, the vertical gradients in all background quantities are
negligible and set them to zero. This will be a good approximation as
long as we consider equilibrium configurations such that
\begin{equation}
\label{eq:logarithmic_derivatives_z}
\left|\frac{d\ln \rho}{d\ln z}  \right|_0  \ll 1  ~, \qquad
\left|\frac{d\ln B_\phi}{d\ln z}\right|_0  \ll 1  ~, \qquad
\textrm{and} \qquad \left|\frac{d\ln \Omega}{d\ln z}\right|_0  \ll 1
~. \qquad
\end{equation}
Note that the solenoidal character of the magnetic field ensures that,
for $\bb{B} = B_{\phi}(r,z) \hat \phi + B_z(r,z) \hat z$, the condition
$\partial B_z/\partial z=0$ holds for arbitrary $z$.

In general, the forces induced by the curvature terms (e.g., the one
proportional to $\epsilon_2$ in  eq.~[\ref{eq:pert_euler_r_grads}])
and those induced by background (logarithmic) gradients in the radial
direction, (e.g., the term proportional to   $d\ln B_\phi/d\ln r|_0$
in the same equation) will not cancel each other.  As a single
exception, for the case in which $B_\phi \propto r^{-1}$, the most
important effects due to the finite curvature of toroidal  field lines
are canceled out by the gradients in the toroidal field.  This
particular case, however, might not be completely relevant to
realistic rotating flows since, in order to ensure force balance when
the thermal pressure can be neglected against magnetic stresses, the
magnetic field strength must decline more slowly than $r^{-1}$ (see,
e.g.,  Kim \& Ostriker 2000). For simplicity, we further focus our
attention on the study of differentially rotating, axisymmetric MHD
flows with locally negligible radial gradients in the background 
density and magnetic field, i.e.,
\begin{equation}
\label{eq:logarithmic_derivatives_r}
\left|\frac{d\ln \rho}{d\ln r}  \right|_0  \ll 1  ~, \qquad
\left|\frac{d\ln B_\phi}{d\ln r}\right|_0  \ll 1  ~, \qquad
\textrm{and} \qquad  \left|\frac{d\ln B_z}{d\ln r}   \right|_0  \ll 1
~. \qquad
\end{equation}
In the rest of the paper, we consider that the only background flow
variable with a non-negligible local radial gradient is the angular
velocity $\Omega \propto r^{-q}$ and set all other radial gradients to
zero.  Note that this assumption is widely invoked in many
investigations of the weak-field MRI (e.g., Blaes \& Balbus 1994;
Balbus \& Hawley 1998; Blaes \& Socrates 2001; Quataert, Dorland, \&
Hammett 2002; Balbus 2003).  The assumption that the only background
variable with a non-negligible  radial gradient is the angular
velocity is also generally a part of the initial set of conditions
used in many numerical analyses of the MRI in the shearing box
approximation (e.g., Hawley, Gammie, \& Balbus 1994, 1995, 1996;
Miller \& Stone 2000).

In spite of being linear in the perturbed quantities, the terms
proportional to $\epsilon_i$  have been neglected in previous local
studies  of the MRI under the assumption that $k_r \gg 1$ and $k_z
\gg1$ (but see also Knobloch 1992 and Kim \& Ostriker 2000).  Although
comparing an imaginary term against a real one in a stability analysis
might seem particularly risky, this might not be a bad argument in
order to neglect the terms proportional to $\epsilon_1$  in equation
(\ref{eq:pert_euler_r_grads})  or $\epsilon_4$ in equations
(\ref{eq:pert_cont_grads}) and (\ref{eq:pert_induc_z_grads}) against
$ik_r$ (but see the discussion in Appendix A). The same could be said
about the terms proportional to $\epsilon_2$ in equation
(\ref{eq:pert_euler_r_grads})  or $\epsilon_3$ in equation
(\ref{eq:pert_euler_phi_grads}) in the limit of a very weak toroidal
component in the magnetic field, given that both of them are
proportional to $v_{{\rm A} \phi}$.  It is not  evident, however, that
we can neglect the terms proportional to either $\epsilon_2$ in
equation  (\ref{eq:pert_euler_r_grads}) or $\epsilon_3$ in equation
(\ref{eq:pert_euler_phi_grads}) if we are to explore the regime of
strong toroidal fields. There are two different reasons for this.  In
order to neglect the term proportional to  $\epsilon_2$ against the
one proportional to $k_r$ in  equation (\ref{eq:pert_euler_r_grads})
we should be able to ensure that the condition
$(\epsilon_2/k_r)(v_{{\rm A}\phi}^2/c_{\rm s}^2) \ll 1$ is always
satisfied, since both terms are proportional to $\delta \rho$.  In
this particular case, neglecting the forces induced by the bending of
toroidal field lines  becomes a progressively worse approximation the
colder the disk is and is not well justified in the limit $c_{\rm s}
\rightarrow 0$.  The case presented in equation
(\ref{eq:pert_euler_phi_grads}) is even harder to justify a priori
since now we would need to guarantee that the  condition
$(\epsilon_3/k_z)(v_{{\rm A} \phi}/v_{{\rm A} z}) (\delta v_{{\rm A}
r} /\delta v_{{\rm A}\phi}) \ll 1$ is always satisfied.  However, this
ratio is not only proportional to  $v_{{\rm A} \phi}/v_{{\rm A} z}$,
which might not be negligible in many astrophysical contexts but,
through the ratio  $\delta v_{{\rm A} r} /\delta v_{{\rm A}\phi}$,  is
also a function of $k_r$, $k_z$, and $\omega(k_r, k_z)$; the magnitude
of this term is therefore  unknown until we solve the problem fully.
A similar situation to this one is encountered if we aim to compare
the term $\epsilon_2  v_{{\rm A} \phi}^2 \delta \rho$  with the term
proportional to $k_z v_{{\rm A} z} \delta v_{{\rm A} r}$ in equation
(\ref{eq:pert_euler_r_grads})   (see \S \ref{subsec:importance} for
further discussion).

For the sake of consistency and in order  not to impose a constraint
on the magnitude of the toroidal \Alfven speed with respect to the
sound speed we keep all the terms proportional to the parameters 
$\epsilon_i$. We will later show that the term proportional to 
$\epsilon_1$ is  negligible when superthermal toroidal fields 
are considered. We will also discuss under which conditions the terms
proportional to $\epsilon_4$ can be neglected and why the terms
proportional to $\epsilon_2$ and $\epsilon_3$ are particularly
important.

\subsection{Dispersion Relation}
\label{sec:dispersion relation}

In order to seek for non-trivial solutions of the homogeneous system
of linear equations
(\ref{eq:pert_cont_grads})-(\ref{eq:pert_induc_z_grads}) we set its
determinant to zero. The resulting characteristic polynomial is
\begin{eqnarray}
\label{eq:disp_full_nodim_kr}
\omega^6  &-& \{(k_z^2+k_r^2)(c_{\rm s}^2+v_{{\rm A} \phi}^2 + v_{{\rm
A}z}^2)  - ik_r [(2\epsilon_1 -\epsilon_2) v_{{\rm A} \phi}^2   +
\epsilon_4 (c_{\rm s}^2 + v_{{\rm   A}z}^2)] + k_z^2 v_{{\rm  A}z}^2 +
\kappa^2 + \epsilon_2 \epsilon_4 v_{{\rm A} \phi}^2 \}  \omega^4 \nonumber \\
&-& \left( 2\epsilon_1 + \epsilon_3 \right) 2 k_z v_{{\rm A}\phi}
v_{{\rm A}z}  \omega^3 + \bigg \lbrace k_z^2 v_{{\rm A}z}^2 [
(k_z^2+k_r^2-\epsilon_4ik_r)  (2c_{\rm s}^2 + v_{{\rm A}\phi}^2 +
v_{{\rm A}z}^2) + ik_r (\epsilon_2-\epsilon_3) v_{{\rm A}\phi}^2 ]
\nonumber \\  &+& k_z^2\left[\kappa^2(c_{\rm s}^2 + v_{{\rm
A}\phi}^2)   + 2 \epsilon_1 \epsilon_4  c_{\rm s}^2 v_{{\rm A}\phi}^2  
+ \epsilon_2 \epsilon_4  v_{{\rm A}\phi}^4  + 2 \frac{d \ln
\Omega}{d \ln r} v_{{\rm A}z}^2 + (\epsilon_2 \epsilon_4   -  2
\epsilon_1 \epsilon_3) v_{{\rm A}\phi}^2 v_{{\rm A}z}^2 \right]
\bigg \rbrace \omega^2 \nonumber \\ &+& 2 k_z^3 v_{{\rm A}\phi}
v_{{\rm A}z} \left[  ( 2 \epsilon_1 + \epsilon_3 + \epsilon_4)
c_{\rm s}^2+ ( \epsilon_2 + \epsilon_3 ) v_{{\rm A}\phi}^2\right]
\omega \nonumber  \\  &-& k_z^4 v_{{\rm A}z}^2 \left[
(k_z^2+k_r^2-\epsilon_4 ik_r) c_{\rm s}^2 v_{{\rm A}z}^2 + 2
\frac{d \ln \Omega}{d \ln r} c_{\rm s}^2 - 2 \epsilon_1 \epsilon_3
c_{\rm s}^2 v_{{\rm A}\phi}^2 - \epsilon_2 \epsilon_3 v_{{\rm
A}\phi}^4 \right] =0 ~,
\end{eqnarray} 
where we have dropped the subscript ``$0$'' in the radial logarithmic
derivative of the angular frequency.  This is the most general
dispersion relation under our  current set of assumptions.  When all
the parameters $\epsilon_i$ are set equal to zero, we recover the results of
previous analyses where the curvature of the toroidal field lines  was
not considered (e.g., Blaes \& Balbus 1994; Balbus \& Hawley 1998),
while when they are set equal to unity we obtain our full dispersion relation.

Although the original linear system
(\ref{eq:pert_cont_grads})-(\ref{eq:pert_induc_z_grads}) related seven
variables (recall that we had eliminated $\delta P$ in terms of
$\delta \rho$ using eq.~[\ref{eq:energy}] which is time-independent),
the characteristic polynomial is only of $6^{{\rm th}}$ degree. This
is easily understood by noting that  equations
(\ref{eq:pert_induc_r_grads}) and (\ref{eq:pert_induc_z_grads}) can be
combined into one single equation expressing the solenoidal character
of the perturbations in the magnetic field, $\del\bcdot \delta \bb{B}
=0$. This implies a relationship between $\delta B_r$ and $\delta B_z$
(or equivalently between $\delta v_{{\rm A} r}$ and $\delta v_{{\rm A}
z}$) that must be satisfied at all times and is, therefore,
independent of $\omega$. The fact that the dispersion relation
(\ref{eq:disp_full_nodim_kr}) is of $6^{th}$ and not of $4^{{\rm th}}$
degree is because  we are taking
into account the effects of finite compressibility.  This can be seen
immediately by taking the limit $c_{\rm s} \rightarrow \infty$.

Once all the
dimensionless variables have been properly defined, it is not evident
that the magnetic-tension terms, proportional to $\epsilon_1$,
$\epsilon_2$, and  $\epsilon_3$, will play a negligible role in
determining the eigenfrequencies $\omega$.  This is because the
non-vanishing toroidal component of the magnetic field introduces odd
powers in the dispersion relation and hence break its even
symmetry. In fact, small modifications in the odd-power coefficients
can and do have an important impact on the nature (real vs. complex)
of the solutions.  As we will see in \S \ref{sec:numerical solutions}
and describe in further detail in   \S \ref{sec:onset of
instabilities},  these curvature terms introduce further coupling
between the radial and toroidal directions, which in turn result in a
strong coupling between the \Alfven and the slow mode.

Also important is the fact that some of the coefficients in the
dispersion relation (\ref{eq:disp_full_nodim_kr})  are no longer real
due to the  factors $ik_r$. The presence of these  terms  does not
allow us to affirm that complex roots will appear in conjugate pairs.
As we discuss in Appendix A, the terms proportional to $ik_r$ play an
important role in determining the stability of modes for which the
ratio $k_r/k_z$ is non-negligible, even in the local limit, i.e.,
when $k_r \gg 1$.  Of course the smaller the ratio $k_r/k_z$, the
smaller the effects  of the factors $ik_r$ will be. If we consider the
limit $k_z \gg k_r$ in equation
(\ref{eq:disp_full_nodim_kr}), 
the imaginary part of all the coefficients in the dispersion
relation will become negligible. 
In this limiting case, whenever a
given complex root is a solution of the dispersion relation
(\ref{eq:disp_full_nodim_kr}) so is its complex conjugate, for the
dispersion relation has real coefficients (see Appendix A).

In the next section, we will show that the dispersion relation
(\ref{eq:disp_full_nodim_kr}) reduces to the  dispersion relations
previously derived in many local studies in different regimes.  It is
important to emphasize, however, that this dispersion relation fully
considers the effects of compressibility and magnetic tension simultaneously
without imposing any restrictions on the field strength or geometry.
This feature is crucial in determining the stability properties of the MHD
flow when strong toroidal fields are considered.

\subsection{Previous Treatments}
\label{sec:previous treatments}
There has been some discussion in the past about the importance of the
curvature terms for the stability of magnetized Keplerian flows
\citep{Knobloch92, GB94}.  In studies in which these terms were
considered \citep{Knobloch92, DK93}, compressibility effects were
neglected. On the other hand, there have also been treatments in which
compressibility was addressed but the curvature terms were neglected
\citep{BB94}.  Both types of studies provided arguments for and
against the importance of these terms. 
The limit of cold MHD flows has been addressed by Kim \& Ostriker (2000).
These authors concluded that when the magnetic field strength  is
superthermal, the inclusion of toroidal fields tends to suppress the
growth of the MRI and that for quasi-toroidal field configurations 
no axisymmetric MRI takes place in the limit $c_{\rm s} \rightarrow 0$. 

Because of the generality of
our treatment, in which both curvature terms and compressibility
effects are fully taken into account, we are able to address all of these
issues in \S \ref{sec:discussion}. For the time being, and as a check, 
we can take the appropriate limits in the general
dispersion relation (\ref{eq:disp_full_nodim_kr}) to recover the
dispersion relations derived in the aforementioned works. 

\emph{Compressibility with no field curvature ---\/}
Setting $\epsilon_i = 0$, for $i=1,2,3,4$,  
and considering perturbations propagating only in the vertical
direction (this can be formally done by taking the limit $k_z \gg k_r$
in equation
[\ref{eq:disp_full_nodim_kr}]) we recover the dispersion relation derived 
in the compressible, weak-field limit by  \citet{BB94},
\begin{eqnarray}
\label{eq:disp_blaes}
&& \omega^6 - [k_z^2(c_{\rm s}^2+v_{{\rm A}\phi}^2+2v_{{\rm A}z}^2) +
\kappa^2] \omega^4 \nonumber \\ &+& k_z^2\left[ k_z^2 v_{{\rm A}z}^2 (2c_{\rm
s}^2 + v_{{\rm A}\phi}^2 + v_{{\rm A}z}^2) + \kappa^2(c_{\rm
s}^2 + v_{{\rm A}\phi}^2) + 2 \frac{d\ln \Omega}{d \ln r} v_{{\rm
A}z}^2 \right]  \omega^2 \nonumber \\ &-& k_z^4 v_{{\rm A}z}^2
c_{\rm s}^2 \left( k_z^2 v_{{\rm A}z}^2 + 2 \frac{d \ln \Omega}{d \ln r}
\right) = 0 ~.
\end{eqnarray}
The stability criterion derived from this dispersion relation is not
different from the one derived, within the Boussinesq approximation, 
by \citet{BH91}. All the perturbations with vertical wavenumber smaller than 
the critical wavenumber $k_{\rm BH}$ are unstable, with 
\begin{eqnarray}
\label{eq:k_BH}
k_{BH}^2 v_{{\rm A}z}^2 \equiv  - 2 \frac{d \ln \Omega}{d \ln r} ~.
\end{eqnarray}
In this case, the strength of the toroidal component of the magnetic
field does not play any role in deciding which modes are subject to
instabilities.

\emph{Field curvature with no compressibility ---\/}
It is important to stress that even in the incompressible limit
not all the terms proportional to $\epsilon_i$ in the dispersion relation
(\ref{eq:disp_full_nodim_kr}) are negligible (of course, the ones
proportional to $\epsilon_2$ are).  
To see that this is the case, we can take the limit 
$c_{\rm s} \rightarrow \infty$ in  the dispersion relation 
(\ref{eq:disp_full_nodim_kr}) to obtain
\begin{eqnarray}
\label{eq:disp_incomp_nodim}
(k_z^2+k_r^2- ik_r \epsilon_4)  \omega^4 
-  k_z^2  [ 2 v_{{\rm A}z}^2 (k_z^2+k_r^2-\epsilon_4ik_r) 
+  \kappa^2  + 2 \epsilon_1 \epsilon_4  v_{{\rm A}\phi}^2 ] \omega^2
- 2 k_z^3 v_{{\rm A}\phi} v_{{\rm A}z} (2 \epsilon_1 + \epsilon_3 +
\epsilon_4) \omega   \nonumber  \\  
+ k_z^4 v_{{\rm A}z}^2 \left[
  (k_z^2+k_r^2-\epsilon_4 ik_r) v_{{\rm A}z}^2 + 2  \frac{d \ln \Omega}{d \ln r} -
  2 \epsilon_1 \epsilon_3 v_{{\rm A}\phi}^2 \right] =0 ~,   %\nonumber  \\  
\end{eqnarray} 
where  we have explicitly left the factors $\epsilon_i$ that should be
considered as unity. This incompressible version of our dispersion
relation is to be compared with the one obtained by \citet{DK93} as
the local limit of the corresponding eigenvalue problem.  Note that, 
in order to compare expression (\ref{eq:disp_incomp_nodim}) with the dispersion
relation (eq.~[37]) presented in \citet{DK93}, it is necessary to 
considered the limit $\partial v_{{\rm A}\phi}/\partial r,  \partial v_{{\rm A}
z}/\partial r \rightarrow 0$ in their equation (9).  We also note that
the radial wavenumber $k_r$ appears in equation
(\ref{eq:disp_incomp_nodim}) only in the combination
$k_z^2+k_r^2-ik_r$ while in equation (37) in \citet{DK93} we only find
it as  $k_z^2+k_r^2$ (i.e., $n^2+k^2$ in their notation). This is
because when taking the local limit, $k_r \gg 1 $, in the process of
deriving their equation (37) from  their equation (9),  the terms
proportional to $ik_r$ were neglected against $k_r^2$ by \citet{DK93}.

When the toroidal component of the magnetic field is negligible, i.e.,
when $v_{{\rm A} \phi} \rightarrow 0$ in equation
(\ref{eq:disp_incomp_nodim}), and we consider vertical modes ($k_z \gg
k_r$), we recover the dispersion relation for the incompressible MRI;
the onset of unstable modes is still given by expression
(\ref{eq:k_BH}).  For weak toroidal fields, i.e., when $v_{{\rm A} \phi}
\ll 1$, we can read off the small corrections to the critical
wavenumber from the constant coefficient,
\begin{eqnarray}
\label{eq:k_c}
(k_z^{0i})^2 v_{{\rm A}z}^2 =  - 2 \frac{d \ln \Omega}{d \ln r} + 2
\epsilon_1 \epsilon_2 v^2_{A \phi}  ~.
\end{eqnarray}
For stronger fields, however, the $\omega = 0$ mode is no longer
unstable (see the Appendix B for a general discussion on the
stability of the $\omega=0$ mode when compressibility and curvature
terms are considered) and it is necessary to solve equation
(\ref{eq:disp_incomp_nodim}) in order to find the critical wavenumber
for the onset of the instability. Roughly speaking, we would expect
the solutions of equation (\ref{eq:disp_incomp_nodim}) to depart
significantly from the solutions to the incompressible version of
the dispersion relation (\ref{eq:disp_blaes}) 
when $v^2_{{\rm A} \phi} \gtrsim  |d\ln \Omega/d
\ln r|$, or $v_{{\rm A} \phi} \gtrsim 1.2$ for a Keplerian disk.
Since in this paper we consider rotationally supported configurations (i.e.,
$v_{{\rm A} \phi} \lesssim 1 $), we will not address the
modifications to the mode structure caused by curvature terms in
incompressible MHD flows.

It is important to stress that, for both dispersion relations
(\ref{eq:disp_blaes}) and (\ref{eq:disp_incomp_nodim}), in the case of
rotationally supported disks, the stability criterion is insensitive
(or, at most, very weakly sensitive, in the incompressible case) to the
magnitude of the toroidal component of the field. As we will see
throughout our study, the stability criteria that emerge from equation
(\ref{eq:disp_full_nodim_kr}) are significantly different from the ones
discussed in this section, when we consider fields for which
$v_{{\rm A} \phi} > c_{\rm s}$. We will also see that the term
proportional to $\epsilon_2$, which depends on curvature and
compressibility effects and is, therefore, absent from either equation
(\ref{eq:disp_blaes}) or (\ref{eq:disp_incomp_nodim}), plays an
important role in determining the mode structure in the general case.

\emph{Cold limit with no field curvature ---\/} Another limit of
interest is the one corresponding to the cold, MHD, cylindrical
shearing flows usually involved in the modeling of cold disk winds
(i.e., far away from the disk). In this context, Kim \& Ostriker
(2000) addressed the behavior of the compressible axisymmetric MRI in
the limit $c_{\rm s} \rightarrow 0$.  These authors obtained a 
dispersion relation considering  both vertical and radial wavenumbers
and derived the criterion for instability associated with it. Their
dispersion relation in the  fully compressible case [eq.~(57)] reads
\begin{eqnarray}
\label{eq:disp_cseq0}
&& \omega^6 - [(k_z^2 + k_r^2)(c_{\rm s}^2  + v_{{\rm
A}\phi}^2+v_{{\rm A}z}^2) + k_z^2 v_{{\rm A}z}^2 + \kappa^2] \omega^4
\nonumber \\  &+& k_z^2 \left[(k_z^2 + k_r^2) v_{{\rm A}z}^2 (2c_{\rm
s}^2  + v_{{\rm A}\phi}^2 + v_{{\rm A}z}^2)  + \kappa^2(c_{\rm s}^2  +
v_{{\rm A}\phi}^2)  + 2 \frac{d\ln \Omega}{d \ln r} v_{{\rm A}z}^2
\right]  \omega^2 \nonumber \\  &-& k_z^4 v_{{\rm A}z}^2 c_{\rm s}^2
\left[ (k_z^2+k_r^2)  v_{{\rm A}z}^2 + 2 \frac{d \ln \Omega}{d \ln r}
\right] = 0 ~.
\end{eqnarray}
This dispersion relation can be obtained from equation
(\ref{eq:disp_full_nodim_kr}) if we set  $\epsilon_i = 0$, for
$i=1,2,3,4$. Note that if we take the limit  $k_z \gg k_r$  in equation
(\ref{eq:disp_cseq0}) we recover equation (\ref{eq:disp_blaes}).

For extremely cold flows we can take the limit $c_{\rm s} \rightarrow
0$ in equation (\ref{eq:disp_cseq0}) to obtain
\begin{eqnarray}
\label{eq:disp_cseq0_cold}
\omega^4 - [(k_z^2+k_r^2)(v_{{\rm A}\phi}^2+ v_{{\rm A}z}^2) + 
  k_z^2 v_{{\rm A}z}^2  + \kappa^2 ] \omega^2  + k_z^2 v_{{\rm A}z}^2
  (k_z^2+k_r^2)(v_{{\rm A}\phi}^2+v_{{\rm A}z}^2)  + \kappa^2 k_z^2
  v_{{\rm A}\phi}^2 + 2 k_z^2  v_{{\rm A}z}^2 \frac{d\ln \Omega}{d \ln
  r} =0 ~. \nonumber \\
\end{eqnarray}
On the other hand, taking the limit $c_{\rm s} \rightarrow 0$ in
equation (\ref{eq:disp_full_nodim_kr}), we obtain the more general dispersion
relation
\begin{eqnarray}
\label{eq:disp_cold_nodim}
\omega^6  &-& \{(k_z^2+k_r^2)(v_{{\rm A} \phi}^2 + v_{{\rm   A}z}^2)
- ik_r [(2\epsilon_1 -\epsilon_2) v_{{\rm A} \phi}^2   + \epsilon_4
v_{{\rm   A}z}^2] + k_z^2 v_{{\rm  A}z}^2 +  \kappa^2 + \epsilon_2
\epsilon_4 v_{{\rm A} \phi}^2 \}  \omega^4 \nonumber \\  &-& \left(
2\epsilon_1 + \epsilon_3 \right) 2 k_z v_{{\rm A}\phi} v_{{\rm A}z}
\omega^3 + \bigg \lbrace k_z^2 v_{{\rm A}z}^2 [
(k_z^2+k_r^2-\epsilon_4ik_r)   ( v_{{\rm A}\phi}^2 + v_{{\rm A}z}^2) +
ik_r (\epsilon_2-\epsilon_3) v_{{\rm A}\phi}^2 ]  \nonumber \\  &+&
k_z^2\left[\kappa^2 v_{{\rm A}\phi}^2   +  \epsilon_2 \epsilon_4
v_{{\rm A}\phi}^4  + 2 \frac{d \ln \Omega}{d \ln r} v_{{\rm A}z}^2 +
(\epsilon_2 \epsilon_4   -  2 \epsilon_1 \epsilon_3) v_{{\rm A}\phi}^2
v_{{\rm A}z}^2 \right]  \bigg \rbrace \omega^2 \nonumber \\ &+& 2
k_z^3 v_{{\rm A}\phi} v_{{\rm A}z} ( \epsilon_2 + \epsilon_3 ) v_{{\rm
A}\phi}^2 \omega + k_z^4 v_{{\rm A}z}^2 \epsilon_2 \epsilon_3 v_{{\rm
A}\phi}^4 =0 ~.
\end{eqnarray} 
Note that in this expression, as it was also the case in the
incompressible limit, several of the terms that are due to the finite
curvature of the toroidal field lines are still present.

Analyzing the limit  $c_{\rm s} \rightarrow 0$  in the  dispersion
relation (\ref{eq:disp_cseq0}),  Kim \& Ostriker (2000) concluded
that toroidal fields tend to suppress the growth of the MRI and that,
for a Keplerian rotation law,  no axisymmetric MRI occurs if
$i<30\degr$, where $i$ is the local pitch angle of the magnetic fields
defined by $i \equiv \tan^{-1} (v_{{\rm A} z}/v_{{\rm A} \phi})$.  
However, the eigenfrequencies satisfying the dispersion relations 
(\ref{eq:disp_full_nodim_kr}) and (\ref{eq:disp_cseq0}) 
in the limit $c_{\rm s} \rightarrow 0$  
are different and so are the criteria
for instability  which they are subject to. In \S
\ref{subsec:Comparison to previous analytical studies}, we comment in
more detail on how the solutions to the dispersion relations
(\ref{eq:disp_full_nodim_kr}) and (\ref{eq:disp_cseq0}) 
differ in the limit  $c_{\rm s} \rightarrow 0$ and on the 
implications  regarding the stabilization of the MRI 
in cold MHD shearing flows.

In order to investigate how previous results from local stability
analyses of the weak field MRI are modified as the strength of the
toroidal field component increases, we will focus our attention on the
stability of modes with $k_z \gg k_r$.\footnote{In Appendix A, we
briefly describe how this results are modified when finite ratios
$k_r/k_z$ are considered.}  This approach is physically motivated,
since vertical modes correspond to the most unstable modes in the well
studied MRI, and is also more tractable mathematically.  In the next
two sections, we will perform a thorough numerical and semi-analytical study
of the general dispersion relation (\ref{eq:disp_full_nodim_kr}) in
the limit  $k_z\gg k_r$, with particular emphasis on the
case of strong toroidal fields.    We will then be in a better
position to understand  the similitudes and differences of our
findings  with those of the aforementioned studies and  we will
address them in \S\ref{sec:discussion}.

\section{\textsc{NUMERICAL SOLUTIONS}}
\label{sec:numerical solutions}

\begin{figure}[tbh]
\begin{center}
\includegraphics[width=\textwidth]{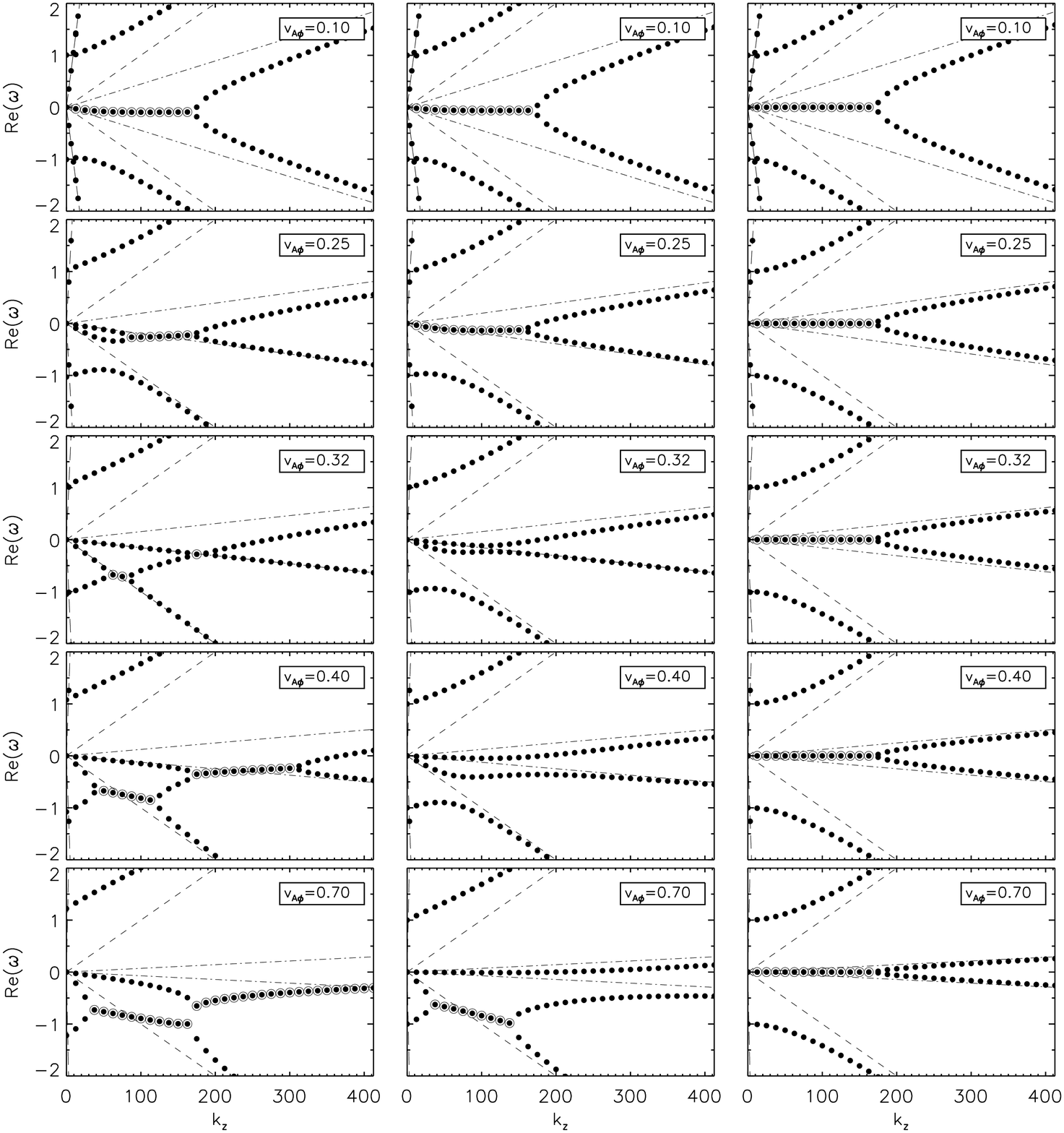}
\caption{The real parts of the numerical solutions to the dispersion
  relation (\ref{eq:disp_full_nodim_kr}) corresponding to a Keplerian
  disk with $c_{\rm s}= 0.05$ and $v_{{\rm A}z}=0.01$.
  \emph{Left panel}: solutions to the full problem
  ($\epsilon_1=\epsilon_2=\epsilon_3=\epsilon_4=1$). \emph{Central
  panel}: the case in which compressibility is neglected in the
  curvature terms ($\epsilon_1=\epsilon_3=\epsilon_4=1$ and
  $\epsilon_2=0$). \emph{Right panel}: the case in which all curvature
  terms are neglected
  ($\epsilon_1=\epsilon_2=\epsilon_3=\epsilon_4=0$). 
  Open circles
  indicate unstable modes (i.e., those with positive imaginary part).
  Long-dashed, short-dashed, and point-dashed lines show the fast,
  \Alfvennospace, and slow modes, respectively, in the limit of no
  rotation.}
\label{fig:comp_real} 
\end{center}
\end{figure}

\begin{figure}[tbh]
\begin{center}
\includegraphics[width=\textwidth]{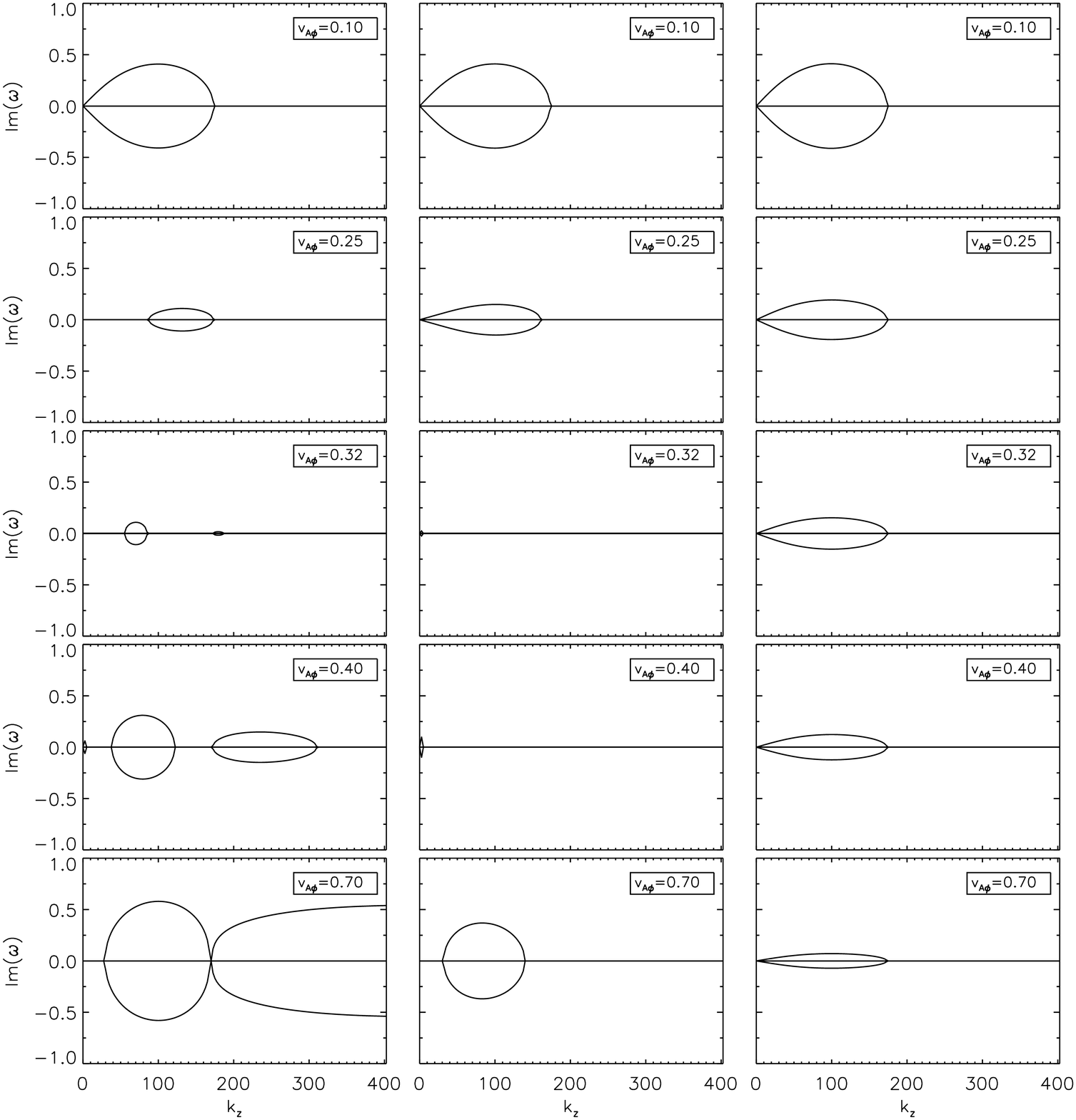}
\caption{The imaginary parts for the cases discussed in Figure
\ref{fig:comp_real}.}
\label{fig:comp_imag} 
\end{center}
\end{figure}

We solved numerically the dispersion relation
(\ref{eq:disp_full_nodim_kr})   for the frequency $\omega$ as a
function of the wavenumber $k_z$, employing Laguerre's root finding
method (Press et al.~1992).   As a typical situation of interest, we
consider a Keplerian disk with  $c_{\rm s}= 0.05$ and $v_{{\rm
A}z}=0.01$.  As it will be seen from the range of values of $k_z$ in
which the various instabilities occur, the case of quasi-toroidal
superthermal fields is perfectly suited to be studied in the local
approximation, i.e., when $k_z \gg 1$, provided that the vertical component of
the magnetic field  is weak enough (i.e., $v_{{\rm A}z} \ll 1$).

To better appreciate the effects that the curvature terms have on the
stability of the modes, a set of solutions to the dispersion relation
(\ref{eq:disp_full_nodim_kr}) is shown in Figures \ref{fig:comp_real}
and \ref{fig:comp_imag}.\footnote{Some animations of the results
presented in Figs. \ref{fig:comp_real}, \ref{fig:comp_imag},
\ref{fig:dlog}, and \ref{fig:sound} are available at
\texttt{http://www.physics.arizona.edu/$\sim$mpessah/research/}}  Each
of the three panels, in both figures, shows the real and imaginary
parts of the solutions for different values of the toroidal field
strength, parameterized by $v_{{\rm A}\phi}$.  The left panel shows
the solutions to the full dispersion relation
(\ref{eq:disp_full_nodim_kr}), i.e., when
$\epsilon_1=\epsilon_2=\epsilon_3=\epsilon_4=1$. The central panel
shows the solutions to equation (\ref{eq:disp_full_nodim_kr}) when
compressibility is neglected in the curvature terms, i.e., when
$\epsilon_1=\epsilon_3=\epsilon_4=1$  and $\epsilon_2=0$.  For the
sake of comparison, the right panel shows the solutions to the
dispersion relation (\ref{eq:disp_blaes}), in which all curvature
terms are neglected.

We first analyze Figure \ref{fig:comp_real}.  When all magnetic
tension terms are neglected (right panel), the qualitative structure
of the normal modes of the plasma is insensitive to the magnitude of
the toroidal field component (see Blaes \& Balbus 1994). However, the
situation is very different when the magnetic tension terms are
included. For weak toroidal fields, i.e., when  $v_{{\rm A}\phi} \lesssim
0.1$, the solutions seem quite insensitive to the curvature terms;
indeed these terms do not seem to play a significant role in altering
the local stability properties of magnetized Keplerian flows compared
to what is quoted elsewhere in the literature.  As we will see
later, for a Keplerian disk, the presence of the curvature terms is
significant once $v_{{\rm A}\phi}^2 \gtrsim c_{\rm s}$, which in this
case translates into $v_{{\rm A}\phi} \gtrsim 0.22$.

For stronger toroidal fields, i.e., when $v_{{\rm A}\phi} \gtrsim
0.2$, the modes with the longest wavelengths become stable when all
curvature terms are included, in sharp contrast to the case in which
$\epsilon_2=0$. For even stronger toroidal fields, i.e., when $v_{{\rm
    A}\phi} \gtrsim 0.3$, a second instability appears at long
wavelengths, while the original instability is suppressed.  When
$v_{{\rm A}\phi} \gtrsim 0.4$, both instabilities coexist as separate
entities and the original instability reaches smaller and smaller
spatial scales, when the magnitude of the toroidal field
increases. For even higher toroidal fields, i.e., when  $v_{{\rm
    A}\phi} \gtrsim 0.7$, the largest unstable wavenumber of the
instability that developed for $v_{{\rm A}\phi} \gtrsim 0.3$
approaches $k_{\rm BH}$ (see eq.~[\ref{eq:k_BH}]).  The major
implication of neglecting compressibility in the curvature  terms is
that the original instability seems to be totally suppressed  for
toroidal fields larger than the ones corresponding to $v_{{\rm A}\phi}
\gtrsim 0.3$.

As it is clear from the dispersion relation
(\ref{eq:disp_full_nodim_kr}),  the presence of the toroidal component
in the field introduces odd powers of the mode frequency $\omega$ and
hence breaks the symmetry between positive and negative real parts of
the solutions. The physical meaning of this is clear.  The phase
velocities of the instabilities are no longer zero and they are
propagating vertically throughout the disk. This, of course, is not
the case for the unstable solutions to the dispersion relation 
(\ref{eq:disp_blaes})
regardless of the magnitude of $v_{{\rm A}\phi}$. In that case, the
most noticeable effect of an increasing toroidal field is to reduce
the phase velocity of the stable modes beyond $k_{\rm BH}$ (which is
itself independent of $v_{{\rm A}\phi}$).

It is also interesting to analyze how the presence of the curvature
terms modifies the growth rates of the unstable modes as a function of
the toroidal magnetic field. This is shown in Figure
\ref{fig:comp_imag}. Again, there are no significant changes for
$v_{{\rm A}\phi} \lesssim 0.1$; however, quite significant
modifications  to the growth rates are present for $v_{{\rm A}\phi}
\gtrsim 0.2$.   The sequence of plots in the left panel shows more
clearly the suppression of the original instability, the appearance of
the instability at low wavenumbers, the return of the instability at
high wavenumbers, and finally the fusion of these last two.  
The right panel in this figure shows the effects that the presence of a strong
toroidal component has on the mode structure when curvature terms are
not considered. In this case, the critical wavenumber for the onset of
instabilities is not modified while there is a clear 
reduction in the growth rate of the non-propagating unstable modes as
the magnitude of the toroidal field component increases.
When the curvature terms are
considered fully, the effects are more dramatic.  Note also that the
growth rate of the original instability is reduced faster from the
first to the second plot in the left panel in Figure
\ref{fig:comp_imag} with respect to their counterparts in the right
panel of the same figure.

\section{\textsc{THE ONSET OF INSTABILITIES}}
\label{sec:onset of instabilities}

\subsection{Unstable Modes}
\label{subsec:unstable modes}

\begin{figure}[!tbh]
\begin{center}
\includegraphics[width=.75\textwidth]{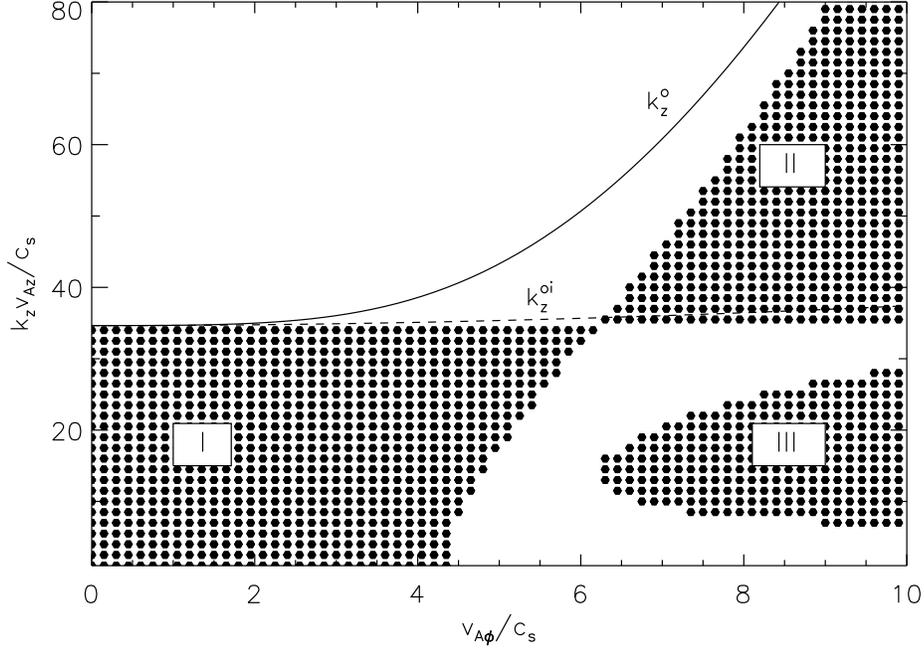}
\caption{The black dots represent unstable modes obtained from solving the
        dispersion relation (\ref{eq:disp_full_nodim_kr}) numerically for
        a Keplerian disk with $c_{\rm s}=0.05$ and $v_{{\rm A} z}=
        0.01$. The solid ($k_z^0$) and dashed ($k_z^{0i}$)  lines correspond
        to the critical wavenumber for which the
        $\omega = 0$ mode exists in the case of a compressible (see
        Appendix B) and an incompressible (discussed in \S
        \ref{sec:mhd equations and dispersion relation}) flow,
        respectively. For strong toroidal fields, compressibility
        plays a crucial role in the stability of the $\omega = 0$
        mode. Note that, in the limit of small $v_{{\rm A}\phi}/c_{\rm
        s}$ we have $k_z^0, k_z^{0i} \rightarrow k_{\rm BH}$, and the trivial
        mode becomes unstable.}
\label{fig:unstable_regions_num}
\end{center}
\end{figure}

In \S \ref{sec:numerical solutions} we presented how the structure of
the various modes evolves as a function of the toroidal field strength
and noted that, for a range of field strengths,  two different
instabilities are clearly  distinguishable.  Here, we obtain the
conditions (i.e., the range of wavenumbers and toroidal field
strengths) for which these unstable modes are present. We start by plotting
in Figure \ref{fig:unstable_regions_num} the range of unstable
wavenumbers as a function of the toroidal field strength. As a
reference, we have plotted the case for a Keplerian disk.  The black
dots in the diagram represent the unstable vertical wavenumbers, in
units of  $k_z v_{{\rm A} z}/c_{\rm s}$, for a given toroidal \Alfven
speed, in terms of $v_{{\rm A}\phi}/c_{\rm s}$.  Three regions of
unstable modes are clearly distinguishable:
\begin{itemize}

\item \emph{Region I\/} shows the evolution of the original
instability present in the topmost three plots in the left panel in
Figure \ref{fig:comp_real}.  This is the region where the MRI
lives. Strictly speaking, the MRI is confined to the region where
$v_{{\rm A} \phi}/c_{\rm s} \ll 1$. As we will comment in \S
\ref{subsec:analytic approximations}, instability I is no longer
incompressible beyond this point.  The maximum wavenumber for which
this instability exists is independent of $v_{{\rm A}\phi}$ and
corresponds to the critical wavenumber for the onset of the MRI (i.e.,
$k_{\rm BH}$ in eq.~[\ref{eq:k_BH}]).   The stabilization of the
long-wavelength perturbations beyond a critical value of the toroidal
\Alfven speed is also evident in this region. For larger toroidal
field strengths, shorter and shorter wavelengths are stabilized up to the ones
corresponding to  $k_{\rm BH}$.

\item  \emph{Region II\/} represents   the evolution of the
instability that is only present for wavenumbers $k_z > k_{ \rm
BH}$. Note that $k_{\rm BH}$ is  now the minimum wavenumber for the
onset of instability II. In this case, increasing $v_{{\rm A}
\phi}/c_{\rm s}$ gives rise to unstable modes with even shorter
wavelengths (two bottommost plots in the left panel of either Figure
\ref{fig:comp_real} or \ref{fig:comp_imag}).

\item \emph{Region III\/} shows the instability  that appears for
intermediate wavenumbers (see for example the third plot in the left
panel in Figure \ref{fig:comp_real}).  Note that the shortest unstable
wavelength in this region approaches $k_{\rm BH}$ for large values of
$v_{{\rm A} \phi}/c_{\rm s}$ (i.e., bottommost plot in the left panel
of either Figure \ref{fig:comp_real} or \ref{fig:comp_imag}).
\end{itemize}

\subsection{Analytic Approximations}
\label{subsec:analytic approximations}

In this section we obtain analytical approximations to  the dispersion
relation (\ref{eq:disp_full_nodim_kr}) in various limits,  which will
help us identify the different critical curves in  Figure
\ref{fig:unstable_regions_num}.

The fast (or magnetosonic) modes are reasonably
well decoupled from the rest of the oscillations (see left panels in
Fig.\ref{fig:comp_real}). By studying the modes that satisfy the
condition $\omega^2 \ll k_z^2 c^2_{\rm s}$, we effectively eliminate
the fast modes from our analysis. This can be done for strong toroidal
fields because, even in the presence of rotation, the magnetosonic
modes are well described by  $\omega^2 \simeq k_z^2 (c^2_{\rm s} +
v^2_{{\rm A}})$. Note that imposing $\omega^2 \ll k_z^2 c^2_{\rm s}$  is
a distinct and weaker condition than asking for the MHD fluid to be
incompressible ($c_{\rm s} \rightarrow \infty$).  By eliminating these
fastest modes, it is possible to find a $4^{\rm th}$ degree dispersion
relation in $\omega$, with solutions that constitute a very good
approximation to the interesting modes seen in Figures
\ref{fig:comp_real} and \ref{fig:comp_imag}.

We first write the equations for the evolution of the perturbations in the
magnetic field. For the sake of clarity, we present the intermediate
steps with the appropriate physical dimensions but we drop the 
index indicating local values. Substituting equations
(\ref{eq:pert_cont_grads}),  (\ref{eq:pert_induc_r_grads}), 
and (\ref{eq:pert_induc_z_grads}) 
in equation (\ref{eq:pert_euler_z_grads}) we obtain $\delta v_z$ 
in terms of $\delta B_\phi$ and $\delta B_z$,
\begin{equation}
\label{eq:vz_Bphi}
\delta v_z = - \frac{k_z \omega c_{\rm s}^2  } {(k_zc_{\rm s})^2 -
\omega^2} \left[ \frac{v_{{\rm A} \phi}^2}{c_{\rm s}^2}\frac{\delta
B_\phi}{ B_\phi} +  \frac{\delta B_z}{B_z} \right] ~.
\end{equation}
Using this result in equation (\ref{eq:pert_induc_phi_grads}) we find,
\begin{equation}
\label{eq:vphi_BrBphi}
i k_z B_z \delta v_\phi = - \frac{d\Omega}{d \ln r} \delta B_r  -i\omega
\delta B_\phi - i \omega B_\phi \frac{(k_zc_{\rm s})^2 } {(k_zc_{\rm s})^2
  - \omega^2} \left[ \frac{v_{{\rm A} \phi}^2}{c_{\rm
      s}^2}\frac{\delta B_\phi}{B_\phi} +  \frac{\delta B_z}{ B_z} \right] ~.
\end{equation}
From  equations (\ref{eq:pert_cont_grads}), (\ref{eq:pert_induc_z_grads}), and
(\ref{eq:vz_Bphi}) we can recast $ \delta \rho$ in terms of $\delta
B_\phi$ and  $\delta B_z$ as
\begin{equation}
\label{eq:rho_Bphi}
\frac{\delta \rho}{\rho} =   - \frac{k_z^2 v_{{\rm A} \phi}^2}{(k_zc_{\rm
    s})^2 - \omega^2} \frac{\delta B_\phi}{ B_\phi} +  \left[ 1 -
  \frac{(k_zc_{\rm s})^2} {(k_zc_{\rm s})^2 - \omega^2} \right]
\frac{\delta B_z}{ B_z} ~.
\end{equation}

Finally, we can write equations (\ref{eq:vz_Bphi})-(\ref{eq:rho_Bphi}) for the
modes with frequencies such that $\omega^2 \ll k_z^2 c^2_{\rm s}$ as,
\begin{equation}
\label{eq:vz_Bphi_limit}
\delta v_z = - \frac{\omega}{k_z}  \left[ \left(\frac{v_{{\rm
A}\phi}}{c_{\rm s}}\right)^2 \frac{\delta B_\phi}{B_\phi} +
i\frac{\epsilon_4}{k_z r}\frac{\delta B_r}{B_z} \right] ~,
\end{equation}
\begin{equation}
\label{eq:vphi_BrBphi_limit}
i k_z B_z \delta v_\phi = - \left[\frac{d\Omega}{d \ln r}  -
  \frac{\epsilon_4}{r} \frac{\omega}{k_z}\frac{B_\phi}{B_z}
  \right]\delta B_r - \left[ 1+  \left( \frac{v_{{\rm A}\phi}}{c_{\rm
      s}}\right)^2 \right] i \omega \delta B_\phi ~,
\end{equation}
and
\begin{equation}
\label{eq:rho_Bphi_limit}
\frac{\delta \rho}{\rho} = -  \frac{v^2_{A \phi}}{c^2_s} \frac{\delta
 B_\phi}{B_\phi} ~,
\end{equation}
where we have used equations (\ref{eq:pert_induc_r_grads}) and
(\ref{eq:pert_induc_z_grads}) to recast $\delta B_z$ in terms of
$\delta B_r$. Note that, neglecting the factor  $\omega^2$ against
$k_z^2 c^2_{\rm s}$ in equation (\ref{eq:vz_Bphi}), and
therefore in equations (\ref{eq:vphi_BrBphi}) and (\ref{eq:rho_Bphi}),
effectively reduces to neglecting the term proportional to $\omega$ in
equation (\ref{eq:pert_euler_z_grads}). Thus, for the modes of
interest, the condition  $\omega^2 \ll k_z^2 c^2_{\rm s}$ is a
statement about force balance in the vertical direction, which is made
explicit in equation  (\ref{eq:rho_Bphi_limit}). In this way,  we can
see how important perturbations in the density are,  in the presence
of strong toroidal fields (see also Balbus \& Hawley 1991).  For
$v_{{\rm A} \phi} \gg c_{\rm s}$,  even small variation in the
toroidal component of the field can have an important impact on the
dynamics of the perturbations.  For this reason, the assumption of an
incompressible MHD flow is not valid, whenever superthermal toroidal
fields are considered. Note that, in order to recover the
incompressible MRI when  $\epsilon_i = 0$, for $i=1,2,3,4$, we have
not neglected the factor unity against $(v_{{\rm A} \phi}/ c_{\rm
s})^2$, in equation (\ref{eq:vphi_BrBphi_limit}).

\begin{figure}[tbh]
\begin{center}
\includegraphics[width=\textwidth]{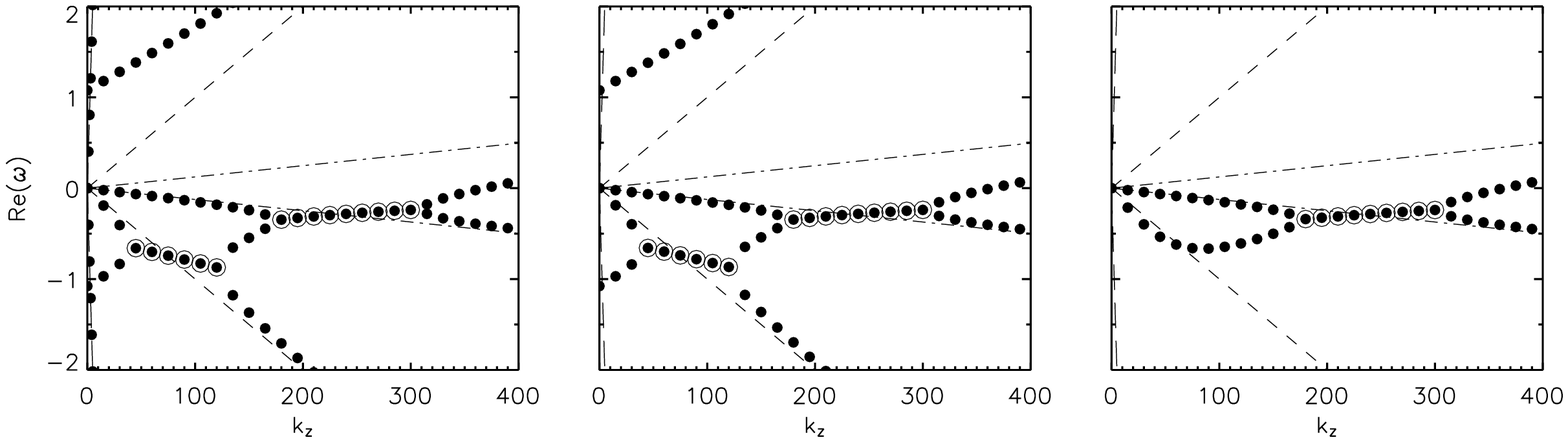}
\includegraphics[width=\textwidth]{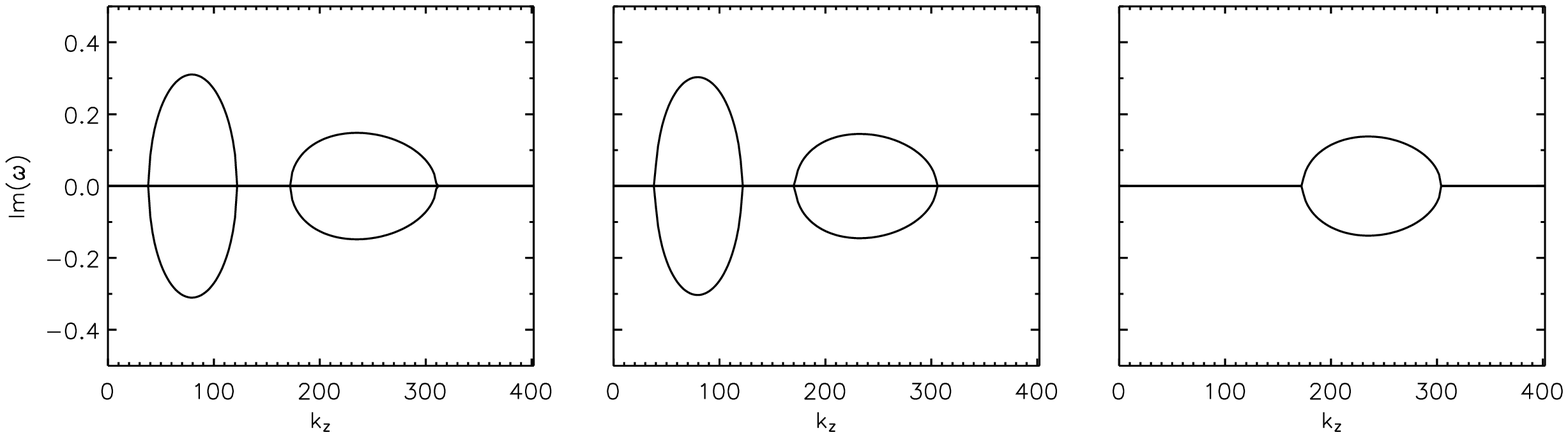}
\caption{\emph{Left panels}: Solutions to the full dispersion relation
  (\ref{eq:disp_full_nodim_kr}). \emph{Central panels}: Solutions to
  the $4^{\rm th}$ order, approximate dispersion relation
  (\ref{eq:disp_large_alpha}), with
  $\epsilon_2=\epsilon_3=\epsilon_4=1$.  \emph{Right panels}:
  Solutions to the $2^{\rm nd}$ order, approximate dispersion relation
  (\ref{eq:disp_2nd}), with $\epsilon_2=\epsilon_3=1$ and
  $\epsilon_4=0$. All solutions correspond to a Keplerian disk with
  $c_{\rm s}= 0.05$, $v_{{\rm A}z}=0.01$, and $v_{{\rm A}
  \phi}=0.40$.  Open circles in upper panels indicate unstable
  modes. Note that the phase velocities of the two instabilities (seen
  in either the leftmost or central upper panels and corresponding to
  Region II and III in \S \ref{subsec:unstable modes}) are similar to
  the phase velocities, positive and negative respectively, of the
  slow mode (point-dashed line) in the limit of no rotation. The fast
  magnetosonic modes can barely be seen close to the left axis in the
  upper left panel.}
\label{fig:approx}
\end{center}
\end{figure}

We now have all the elements to write equations
(\ref{eq:pert_euler_r_grads}) and (\ref{eq:pert_euler_phi_grads}) in
terms of $\delta B_r$ and $\delta B_\phi$. Using equations
(\ref{eq:vz_Bphi_limit})-(\ref{eq:rho_Bphi_limit}), valid in the limit
$\omega^2 \ll k_z^2 c^2_{\rm s}$, we obtain, in terms of the
dimensionless variables,
\begin{eqnarray}
\label{eq:motion_Br}
- \omega^2 \delta B_r + 2i\omega \left[ 1 + \left(\frac{v_{{\rm A}
  \phi}}{c_{\rm s}} \right)^2 \right] \delta B_\phi = &-&  \left[
  2\frac{d \ln \Omega}{d \ln r} + (k_z v_{{\rm A} z})^2   -
  2\epsilon_4 \frac{\omega}{k_z}\frac{v_{{\rm A} \phi}}{v_{{\rm A}
  z}}\right] \delta B_r  \nonumber \\ &-& i k_z v_{{\rm A} \phi}
  v_{{\rm A}z}    \left[ 2\epsilon_1 + \epsilon_2 \left(\frac{v_{{\rm
  A} \phi}}{c_{\rm s}} \right)^2 \right] \delta B_\phi ~,
\end{eqnarray}
\begin{equation}
\label{eq:motion_Bphi}
 - \omega^2 \delta B_\phi \left[ 1 + \left(\frac{v_{{\rm
 A}\phi}}{c_{\rm s}}\right)^2 \right]  - i \omega \left[2 + \epsilon_4
 \frac{\omega}{k_z}\frac{v_{{\rm A} \phi}}{v_{{\rm A} z}} \right]\delta
 B_r = - (k_z v_{{\rm A} z})^2 \delta B_\phi + i k_z v_{{\rm A} \phi}
 v_{{\rm A}z} \epsilon_3 \delta B_r ~.
\end{equation}
These equations are the generalization of the set of equations used to
illustrate the physics behind the weak-field 
MRI as a system of masses coupled by
a spring in a differentially rotating background. Indeed, in the
incompressible limit and neglecting the curvature terms proportional
to $\epsilon_i$, for $i=1,2,3,4$,
we recover the set of equations presented elsewhere
(Balbus \& Hawley 1992, 1998).

\begin{figure}[tbh]
\begin{center}
\includegraphics[width=.75\textwidth]{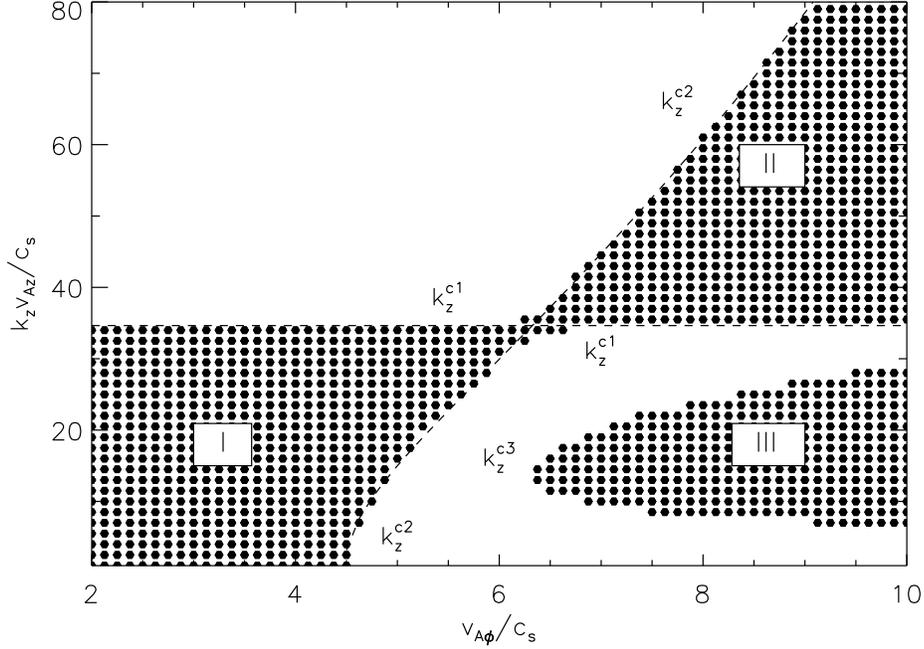}
\caption{The black dots represent unstable modes satisfying the
        approximate instability criteria (\ref{eq:D_4}), described in
        \S \ref{subsec:analytic approximations}.  The dashed lines,
        labeled by $k_z^{{\rm c} 1}$ and $k_z^{{\rm c} 2}$, 
	are the limits of Regions I
        and II obtained analytically, also in \S \ref{subsec:analytic
        approximations}. The onset of instability III is labeled by
        $k_z^{{\rm c} 3}$. As in Figure
        \ref{fig:unstable_regions_num}, we have assumed a
        Keplerian disk with $c_{\rm s}=0.05$ and $v_{{\rm A}z}= 0.01$.}
\label{fig:unstable_regions_anal}
\end{center}
\end{figure}

Setting the determinant of the linear system
(\ref{eq:motion_Br})-(\ref{eq:motion_Bphi}) equal to zero and taking
the limit $v_{{\rm A} \phi}  \gg  c_{\rm s} $ provides the following
approximate dispersion relation that is valid for strong toroidal
fields\footnote{Note that,  had we taken the opposite limit, i.e.,  $
c_{\rm s} \gg v_{{\rm A} \phi}$,  we would have recovered the
dispersion relation (\ref{eq:disp_incomp_nodim}) in the limit 
$k_r/k_z \rightarrow 0$.},
\begin{equation}
\label{eq:disp_large_alpha}
\omega^4 - (\kappa^2 +  k_z^2 v_{{\rm A} z}^2  +\epsilon_2\epsilon_4
 v_{{\rm A} \phi}^2) \omega^2 - 2 k_z v_{{\rm A} \phi} v_{{\rm A} z}
 (\epsilon_2 + \epsilon_3) \omega  + k_z^2 v_{{\rm A} z}^2 \left[
 \frac{c^2_{\rm s}}{v^2_{{\rm A} \phi}} \left( k_z^2 v_{{\rm A} z}^2 + 2
 \frac{d \ln \Omega}{d \ln r}\right) - \epsilon_2 \epsilon_3 v_{{\rm
 A} \phi}^2 \right]=0 ~.
\end{equation}
Note that we have not neglected the factor  $ c^2_{\rm s}/v^2_{{\rm
A} \phi}$ in the last term in equation (\ref{eq:disp_large_alpha})
because its contribution is non-negligible at large wavenumbers.
The solutions to the dispersion relation (\ref{eq:disp_large_alpha}),
for a Keplerian disk with $c_{\rm s}= 0.05$, $v_{{\rm A}z}=0.01$, 
and $v_{{\rm A} \phi}=0.4$, are shown in the central panels in Figure
\ref{fig:approx}. For the sake of comparison, the left panels in the
same figure show the solutions of the full dispersion relation
(\ref{eq:disp_full_nodim_kr}). The solutions to the approximate
dispersion relation (\ref{eq:disp_large_alpha}) 
are in excellent agreement with the solutions to 
the general dispersion relation (\ref{eq:disp_full_nodim_kr})
for which $\omega^2 \ll k_z^2 c^2_{\rm s}$.

Note that the term proportional to $\epsilon_1$ is not present in
equation (\ref{eq:disp_large_alpha}).  This feature has important
consequences for us to understand the physics behind the stability of
strongly magnetized compressible flows.  It has been suggested
\citep{CP95} that the magnetic tension term $ B_\phi \delta B_\phi /
r_0$ (i.e., the one proportional to $\epsilon_1$ in
eq.~[\ref{eq:pert_euler_r_grads}]) is responsible for the
stabilization of long-wavelength perturbations via the restoring
forces provided by strong toroidal field lines in  incompressible MHD
flows.  This argument sounds compelling, but  we can see from the last
term in equation (\ref{eq:motion_Br}) that the term proportional to
$\epsilon_1$  is not dynamically important for compressible flows in
which $v_{{\rm A} \phi} \gg c_{\rm s}$.  At least in the radial
direction, it is rather the term proportional to $\epsilon_2$ the one
governing the deviation compared to the stability properties of weak
toroidal fields. This is in complete agreement with equation
(\ref{eq:rho_Bphi_limit}).

The dispersion relation (\ref{eq:disp_large_alpha}) is of the form
\begin{equation}
\omega^4 + b_2\omega^2 + b_1 \omega  + b_0 = 0 ~.
\end{equation}
For this $4^{\rm th}$ order equation to have complex roots
(corresponding to unstable modes), its discriminant has to be
negative, i.e., 
\begin{equation}
\label{eq:D_4}
D_4(v_{{\rm A}\phi}, k_z v_{{\rm A} z})=  -4 b_2^3 b_1^2 - 27 b_1^4 + 16
b_0 b_2^4 -128 b_2 b_0^2 + 144 b_2 b_1^2 b_0 + 256 b_0^3 < 0 ~.
\end{equation}

The modes satisfying this condition are shown as black dots in Figure
\ref{fig:unstable_regions_anal}. This analytical criterion agrees well
with the numerical results for most of the parameter space
($v_{{\rm A}\phi}/c_{\rm s}$, $k_z v_{{\rm A} z}/c_{\rm s}$) with the
exception of some of the unstable modes close to the separatrix of the
Regions I and II, defined in \S \ref{subsec:unstable modes}.

\emph{Limiting wavenumbers for Regions I and II.---\/} The modes
satisfying the condition $D_4=0$ correspond to the limits of Regions
I, II, and III in Figure \ref{fig:unstable_regions_anal}. Their
analytical expressions, however, are complicated. Yet, some more
progress can be made by realizing that the solutions to the second
order equation obtained by simply dropping the $\omega^4$ term in
equation (\ref{eq:disp_large_alpha}),
\begin{equation}
\label{eq:disp_2nd}
(\kappa^2 +  k_z^2 v_{{\rm A} z}^2 + \epsilon_2\epsilon_4  v_{{\rm A}
 \phi}^2) \omega^2 + 2 k_z v_{{\rm A} \phi} v_{{\rm A} z} (\epsilon_2 +
 \epsilon_3) \omega  - k_z^2 v_{{\rm A} z}^2 \left[ \frac{c^2_{\rm
 s}}{v^2_{{\rm A} \phi}} \left( k_z^2 v_{{\rm A} z}^2 + 2 \frac{d \ln
 \Omega}{d \ln r}\right) - \epsilon_2 \epsilon_3 v_{{\rm A} \phi}^2
 \right]=0 ~, \nonumber \\
\end{equation}
constitute a very good approximation to the solutions of the
dispersion relation (\ref{eq:disp_full_nodim_kr}) whenever the frequencies
of the modes satisfy $\omega^2 \ll 1$.  This can be
appreciated by comparing the left and right panels in Figure
\ref{fig:approx}. The physics behind this approximation is not as
direct as the physics behind the condition $\omega^2 \ll k_z^2 c^2_{\rm
  s}$, but it can also be understood in terms of force
balance, this time in the radial direction. The dispersion relation
(\ref{eq:disp_2nd}) can be obtained by neglecting the term
proportional to $\omega^2$
in equation (\ref{eq:motion_Br}), setting to zero the determinant
of the resulting linear system given by equations 
(\ref{eq:motion_Br})-(\ref{eq:motion_Bphi}) and taking the limit 
$v_{{\rm A}\phi} \gg c_{\rm s}$.
This approximation is equivalent to neglecting the term proportional to
$\omega$ in equation (\ref{eq:pert_euler_r_grads}) and hence related
to neglecting  the radial acceleration experienced by a displaced fluid
element.

Setting the discriminant of equation (\ref{eq:disp_2nd}) to zero,
gives an equation in $k_z$ with solutions that are the limiting wavenumbers
for the onset of instabilities I and II in Figure
\ref{fig:unstable_regions_anal}, i.e.,
\begin{eqnarray}
\label{eq:D_2}
D_2(v_{{\rm A}\phi}, k_z v_{{\rm A} z}) =  (k_z v_{{\rm A} z})^4 &+&
\left[ \kappa^2 + 2\frac{d \ln \Omega} {d \ln r} - v_{{\rm A} \phi}^2
\left( \frac{v_{{\rm A} \phi}^2}{c_{\rm s}^2} - \epsilon_4
\frac{c_{\rm s}^2}{v_{{\rm A} \phi}^2} \right) \right] (k_z v_{{\rm A}
z})^2  \nonumber \\ &+& 2\frac{d \ln \Omega} {d \ln r} \left[\kappa^2
- v_{{\rm A} \phi}^2 \left( \frac{v_{{\rm A} \phi}^2}{c_{\rm s}^2} -
\epsilon_4 \frac{c_{\rm s}^2}{v_{{\rm A} \phi}^2} \right) \right] -
\epsilon_4 v_{{\rm A} \phi}^4 = 0 ~.
\end{eqnarray}
Here, we have set $\epsilon_2=\epsilon_3=1$ but have explicitly left
$\epsilon_4$ to show that its contribution to the onset of
instabilities I and II is not important when $v_{{\rm A}\phi} \gg
c_{\rm s}$, as long as we are considering a rotationally supported
disk. We mention, however, that the numerical solutions  show
that the contribution of the term proportional to  $\epsilon_4$ is
small but not negligible for the unstable modes in region III.
Neglecting the terms proportional to $\epsilon_4$, the solutions to
equation (\ref{eq:D_2}) are simply
\begin{equation}
(k_z^{\rm c} v_{{\rm A} z})^2 = \frac{1}{2} \left[ \frac{v_{{\rm A}
\phi}^4}{c_{\rm s}^2} -  \left(\kappa^2 + 2\frac{d \ln \Omega}{d \ln
r}\right) \right]  \pm  \frac{1}{2} \left| \frac{v_{{\rm A}
\phi}^4}{c_{\rm s}^2} - 4 \right| ~.
\end{equation}
One of these solutions coincides always with $k_{\rm BH}$
(eq.~[\ref{eq:k_BH}]),
\begin{equation}
\label{eq:kc1}
(k_z^{{\rm c} 1} v_{{\rm A} z})^2 = - 2\frac{d \ln \Omega}{d \ln r} ~,
\end{equation}
and the other one is
\begin{equation}
\label{eq:kc2}
(k_z^{{\rm c} 2} v_{{\rm A} z})^2 = \frac{v_{{\rm A} \phi}^4}{c_{\rm s}^2} -
\kappa^2 ~.
\end{equation}

The modes with wavenumbers in the range 
$[min(k_z^{{\rm c} 1},k_z^{{\rm c} 2}),max(k_z^{{\rm c} 1},k_z^{{\rm c} 2})]$ 
are unstable.
In Figure \ref{fig:unstable_regions_anal},
the critical curves $k_z^{{\rm c} 1}(v_{{\rm A} \phi})$ and 
$k_z^{{\rm c} 2}(v_{{\rm A} \phi})$ are shown, with the proper
normalization, as dashed lines.
The critical wavenumber $k_z^{{\rm c} 2}$ in equation (\ref{eq:kc2}) will be
positive only for toroidal \Alfven speeds larger than
\begin{equation}
\label{eq:vphic_limit_I}
v_{{\rm A} \phi}^{{\rm I}} = \sqrt{\kappa  c_{\rm s}} ~.
\end{equation}
This is the critical value of the \Alfven speed beyond which the modes
with longest wavelength in Region I (see
Fig. \ref{fig:unstable_regions_anal}) begin to be stable.  For a
Keplerian disk, the epicyclic frequency coincides with the orbital
frequency and thus, in dimensionless units, $\kappa^2=1$.  In this
case, the critical \Alfven speed for $k_z^{{\rm c} 2}$ to be positive
corresponds to $v_{{\rm A} \phi} = 0.223$. This is the reason for
which the long-wavelength modes are already stable in the second plot
in the left panel in Figure \ref{fig:comp_real}, where $v_{{\rm A}
\phi}=0.25$.

Incidentally, we find that the values of toroidal \Alfven speeds for
which the standard MRI gives the appropriate range of unstable modes
are not as restricted to $v_{{\rm A} \phi} \ll c_{\rm s}$ but rather
to $v_{{\rm A} \phi} \ll  \sqrt{\kappa  c_{\rm s}}$.  For  $v_{{\rm A}
\phi} \gtrsim  \sqrt{\kappa  c_{\rm s}}$, the standard MRI is
stabilized at low wavenumbers.   We point out that, Papaloizou \&
Szuszkiewicz (1992) found, by means of a global stability analysis of a
compressible flow, that for a  slim disk threaded only by a
vertical field, the flow is stable if the vertical \Alfven speed
exceeds, within a factor of order unity, the geometrical mean of the
sound speed and the rotational speed. In dimensionless units, this
stability criterion translates into $v_{{\rm A} z} \gtrsim
\sqrt{c_{\rm s}}$.

The limiting case in which $k_z^{{\rm c} 1}=k_z^{{\rm c} 2}$, is reached for 
\begin{equation}
\label{eq:vphic_limit_II}
v^{{\rm II}}_{{\rm A} \phi} = \sqrt{2 c_{\rm s}} ~. 
\end{equation}
Note that, for $c_{\rm s} =0.05$, this corresponds to a
value for the critical toroidal \Alfven speed of $v_{{\rm A} \phi} =
0.316$. This situation is to be compared with the mode structure in
the third plot in the left panel in Figure \ref{fig:comp_real}, where
$v_{{\rm A} \phi} = 0.32$. 

\emph{Limiting wavenumbers for Region III.---\/} In the previous section
we presented some useful analytical approximations to describe the
dependence of the critical values of the toroidal \Alfven speeds and
wavenumbers defining Regions I and II on the different quantities
characterizing the MHD flow.  We could not, however, find simple 
analytical expressions to describe satisfactorily the corresponding
behavior of the critical values defining Region III. 
We will describe next how the different unstable regions 
in Figure \ref{fig:unstable_regions_num} depend on the
magnitude of the sound speed and the steepness of the rotation
profile.

\section{\textsc{DISCUSSION}}
\label{sec:discussion}

In this section, we address several issues related to the importance
of the curvature terms in determining the stability criteria  obeyed
by the solutions to the dispersion relation
(\ref{eq:disp_full_nodim_kr}).  We comment on some controversies raised
by previous investigations that have treated the standard MRI taking
into account, in various ways, either compressibility or curvature of
the background magnetic field. We also comment on the 
importance in the outcome of the instabilities played  by the magnetic
tension produced by toroidal filed lines in the limit of cold MHD
flows. We highlight the similarities and differences of our findings
with the results of \citet{CP95}, who also found the emergence of a
new (but different) instability for strong toroidal fields in the case
of an incompressible MHD flow.  Finally, we address the potential
implications of our findings for shearing box models in which magnetic tension
terms, induced by the curvature of the background field, are not considered.

\subsection{Importance of Curvature Terms}
\label{subsec:importance}

\begin{figure}[tbh]
\begin{center}
\includegraphics[width=.75\textwidth, angle=90]{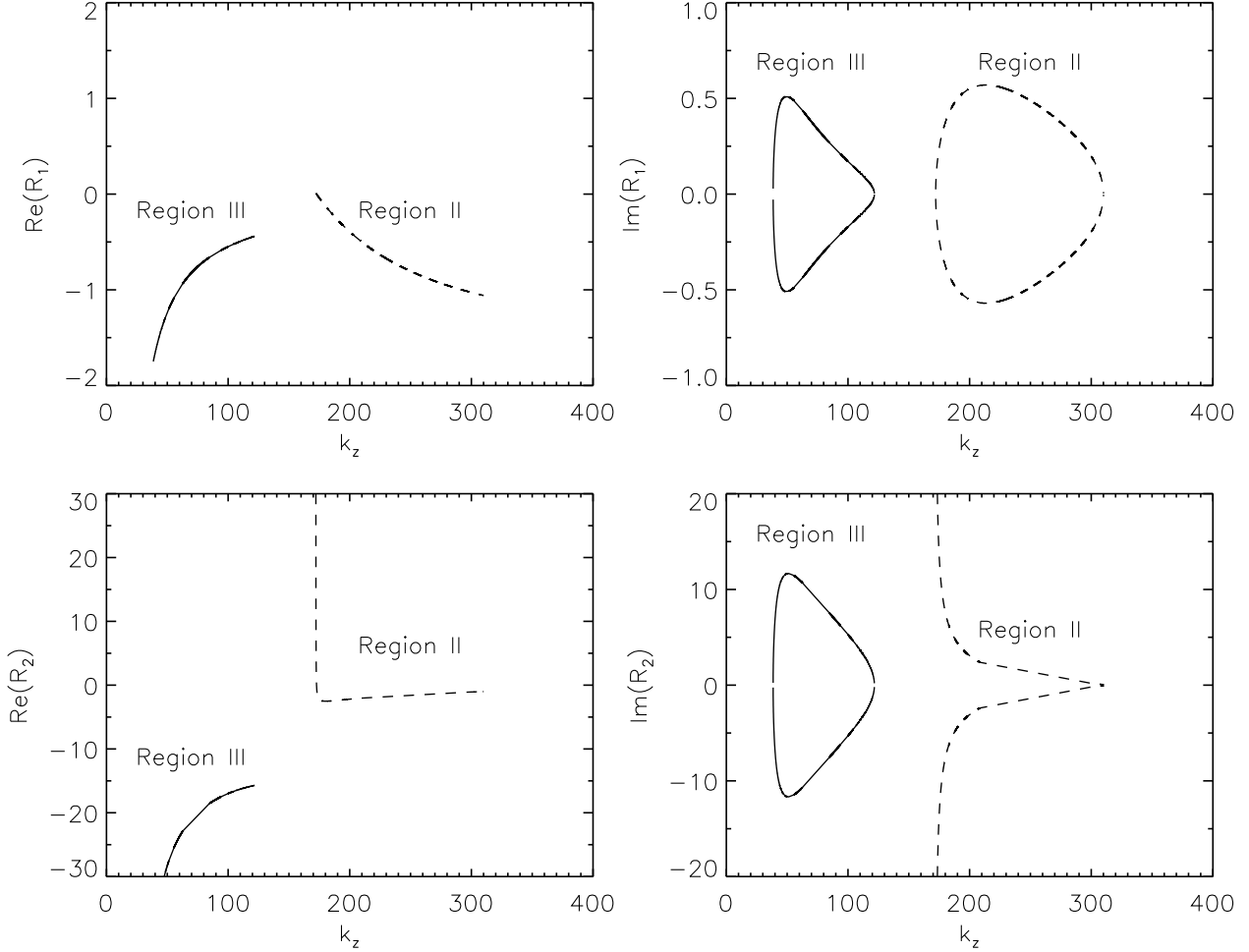}
\caption{The importance of the curvature terms proportional to
  $\epsilon_2$ and $\epsilon_3$, as defined by the ratios $R_1$
  (eq.~[\ref{eq:R1_nodim}]) and $R_2$ (eq.~[\ref{eq:R2_nodim}]).  For
  illustrative purposes, we have considered a Keplerian disk with
  $c_{\rm s}=0.05$, $v_{{\rm A}z}= 0.01$, and $v_{{\rm A} \phi}= 0.4$.}
\label{fig:ratios}
\end{center}
\end{figure}

In section \S\ref{sec:mhd equations and dispersion relation} we
mentioned that the terms proportional to $\epsilon_i$, for
$i=1,2,3,4$, are usually neglected in local stability analyses
due to their $1/r_0$ dependence. Some
of these terms, however, are also proportional to the magnitude of the
toroidal field. In this paper, we found that, when
strong toroidal fields are considered, these terms led to substantial
modifications to the stability criteria of MHD modes known to be valid
in the limit of weak fields.
After solving the full problem, we are in a better
position to understand why this is the case.

To illustrate the point, consider the ratio of the term proportional to $k_z$
to the one proportional to $\epsilon_2$ in equation 
(\ref{eq:pert_euler_r_grads}) and the ratio of the term 
proportional to $k_z$ to the one proportional to $\epsilon_3$ in equation
(\ref{eq:pert_euler_phi_grads}), i.e.,
\begin{equation}
R_1 \equiv \frac{\epsilon_2}{ik_z} \frac{v_{{\rm A} \phi}^2}{v_{{\rm A} z}}
\frac{\delta \rho}{\delta v_{{\rm A} r}}    \qquad {\rm and} \qquad R_2 \equiv
\frac{\epsilon_3}{ik_z} \frac{v_{{\rm A} \phi}}{v_{{\rm A} z}}
\frac{\delta v_{{\rm A} r}}{\delta v_{{\rm A} \phi}}~.
\end{equation}
In order to ensure that the contributions due to curvature are
negligible in a local analysis regardless of the magnitude of the
toroidal field component, we should be able to ensure that the
conditions $R_1 \ll 1$ and  $R_2 \ll 1$ hold in the limit of large $k_z$
for any value of  $v_{{\rm A} \phi} \lesssim 1$.   While it is
encouraging that both dimensionless ratios are proportional to $1/k_z$,
they are also proportional to the ratio of perturbed quantities, which
we do not know \emph{a priori}.   It is only after having found
the eigenfrequencies $\omega(k_z)$ by taking into account all the
curvature terms that we can properly address this issue.

\begin{figure}[tbh]
\begin{center}
\includegraphics[width=.475\textwidth]{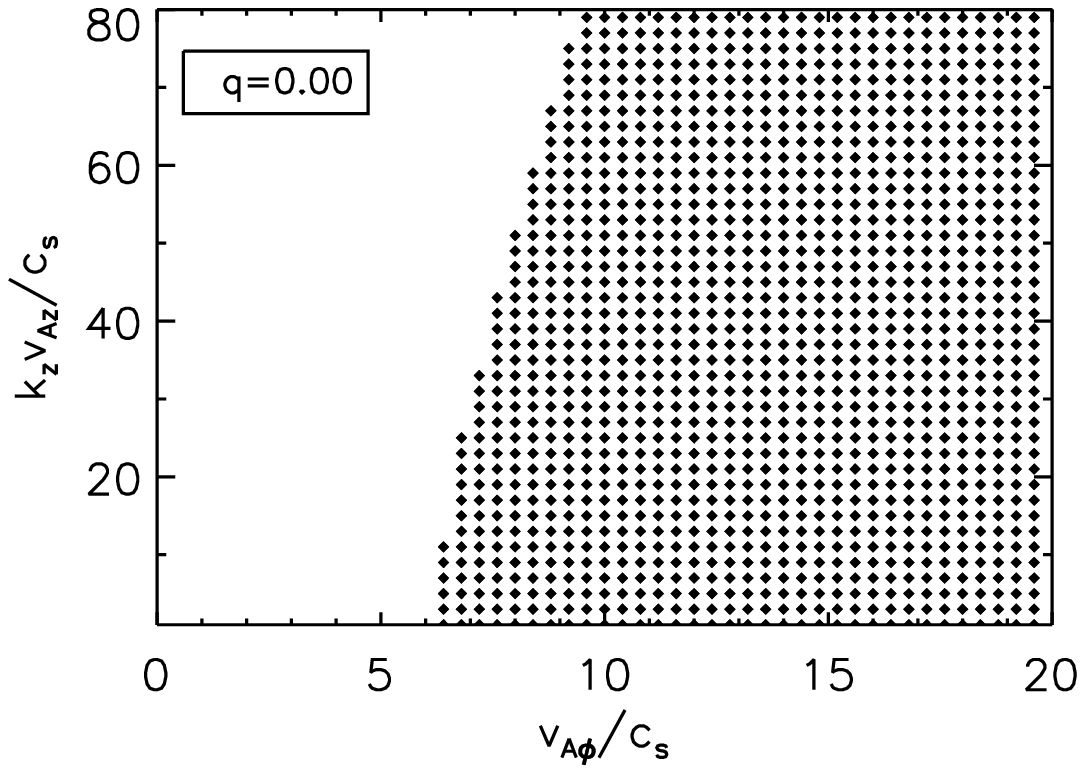}
\includegraphics[width=.475\textwidth]{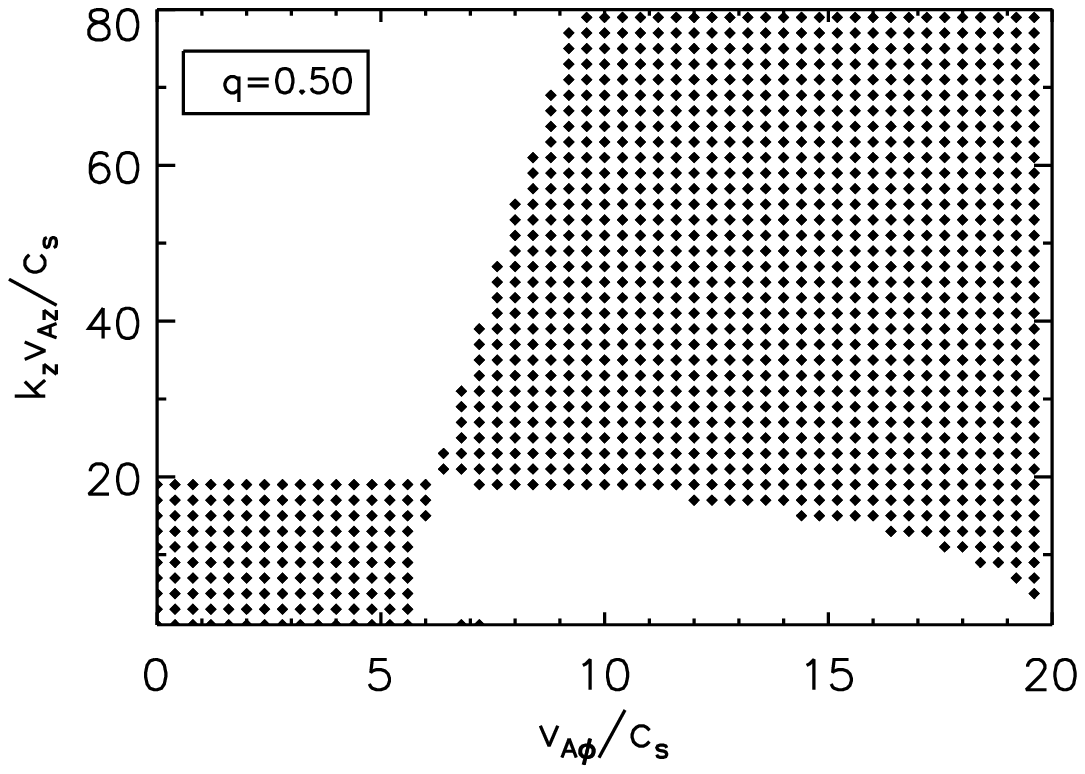}
\includegraphics[width=.475\textwidth]{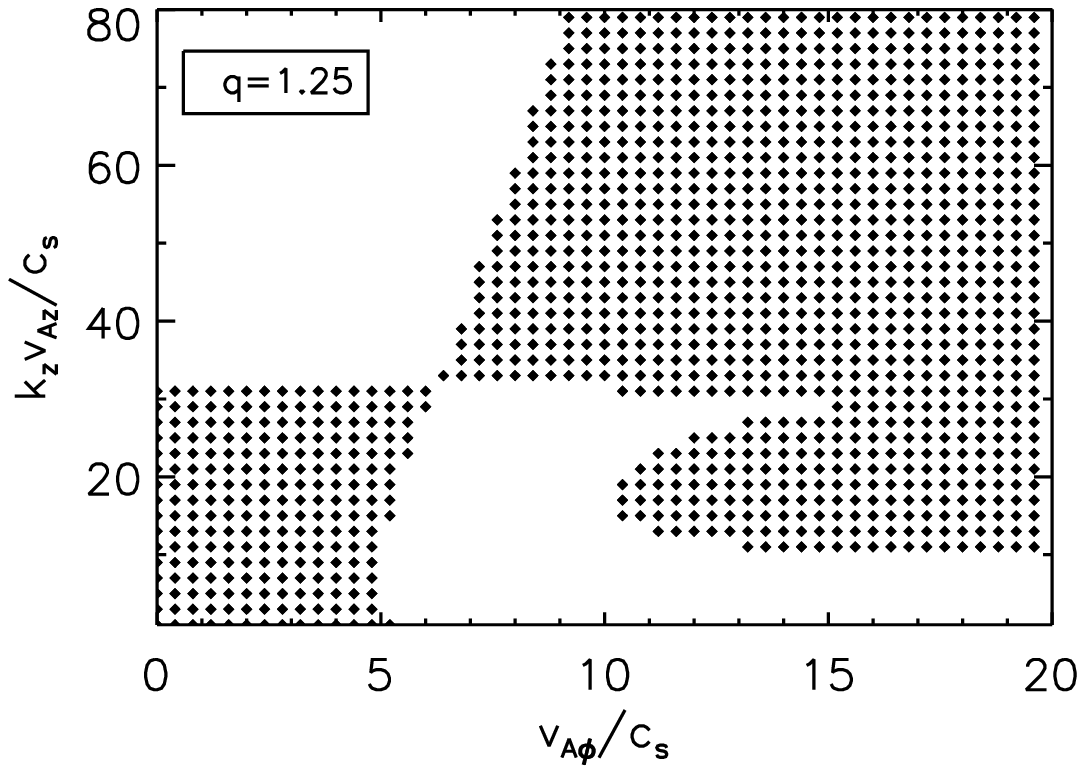}
\includegraphics[width=.475\textwidth]{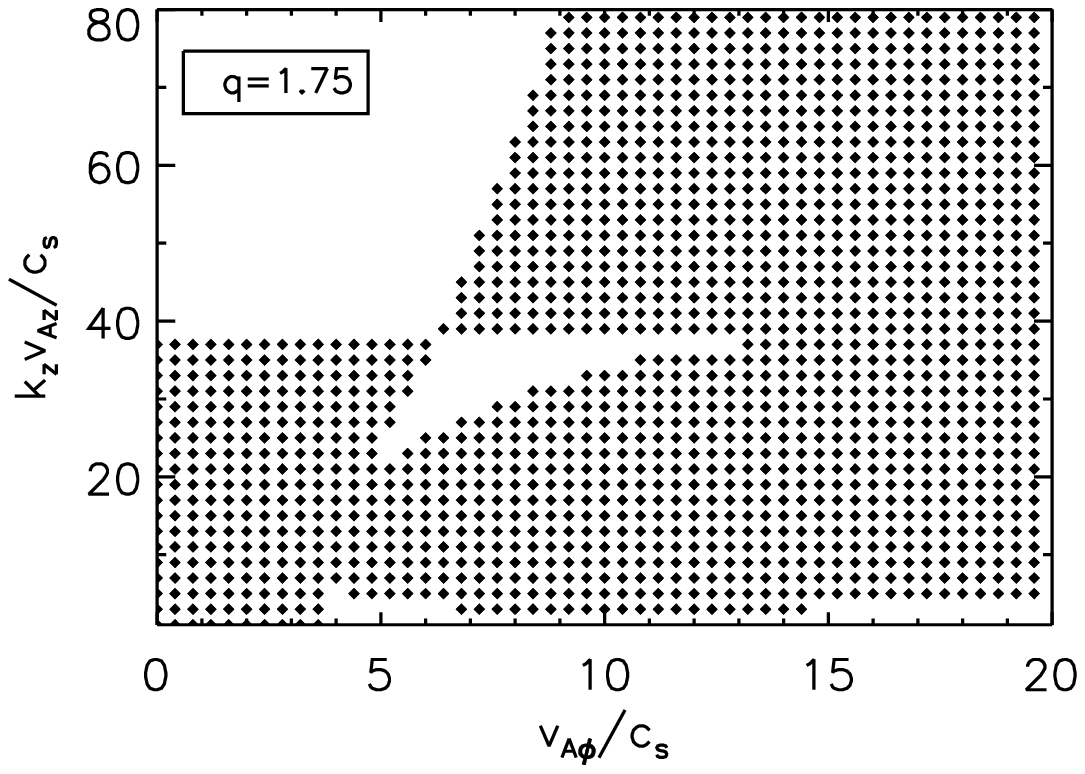}
\caption{The black dots represent unstable modes obtained from solving
        numerically the dispersion relation (\ref{eq:disp_full_nodim_kr})
        as a function of the toroidal \Alfven speed. As an example, we have
        assumed $c_{\rm s}=0.05$ and $v_{{\rm A} z}=
        0.01$. In each plot, we consider different values of the rotational
        profile, $q = - d\ln \Omega/ d\ln r$. Note that,
        the highest value of the local toroidal \Alfven speed considered
        here corresponds to the local circular velocity.}
\label{fig:dlog}
\end{center}
\end{figure}

We can calculate how the ratios $R_1$ and $R_2$ depend on the
wavenumber $k_z$ as follows.
The ratio $R_1$ can be recast using equations
(\ref{eq:rho_Bphi_limit}) and (\ref{eq:motion_Br}) as
\begin{equation}
\label{eq:R1_nodim}
R_1 = \frac{\epsilon_2}{k_z} \frac{v_{{\rm A}\phi}}{v_{{\rm A} z}}
     \left(\frac{v_{{\rm A}\phi}}{c_{\rm s}}\right)^2 
     \frac{ \omega^2 - \left(2 \frac{d\ln \Omega}{d\ln r} + 
     k_z^2 v_{{\rm A} z}^2 -2 \epsilon_4\frac{\omega}{k_z}\frac{v_{{\rm
     A}\phi}}{v_{{\rm A} z}} \right) } 
     {2\omega \left[ 1+\left(\frac{v_{{\rm A}\phi}}{c_{\rm
     s}}\right)^2  \right] + k_z v_{{\rm A} z} v_{{\rm A}\phi}
     \left[ 2\epsilon_1 + \epsilon_2 \left(\frac{v_{{\rm A}\phi}}
     {c_{\rm s}}\right)^2  \right]} ~.
\end{equation}
In a similar way, we can rewrite the ratio $R_2$ using equation
(\ref{eq:motion_Bphi}) as
\begin{equation}
\label{eq:R2_nodim}
R_2 = \frac{\epsilon_3}{k_z} \frac{v_{{\rm A}\phi}}{v_{{\rm A} z}}
\frac{\omega^2 \left[ 1+\left(\frac{v_{{\rm A}\phi}}
{c_{\rm s}}\right)^2  \right] - k_z^2 v^2_{{\rm A} z}}
{\omega \left[ 2+ \epsilon_4  \frac{\omega}{k_z}\frac{v_{{\rm
A}\phi}}{v_{{\rm A} z}} \right] +\epsilon_3 k_z v_{{\rm A} z} 
v_{{\rm A}\phi} } ~.
\end{equation}
For the sake of simplicity, let us consider a given value for the
toroidal \Alfven speed, e.g., $v_{{\rm A}\phi}=0.4$.  Figure
\ref{fig:ratios} shows the dependence of the ratios $R_1$ and $R_2$
on wavenumber for the unstable.  The eigenfrequencies
$\omega(k_z)$ were obtained by solving equation
(\ref{eq:disp_full_nodim_kr}) with $\epsilon_i=1$, for $i=1,2,3,4$,
considering a Keplerian disk with  $c_{\rm s}=0.05$ and $v_{{\rm
A}z}=0.01$. The ratios $R_1$ and $R_2$ for the unstable modes (with
$v_{{\rm A}\phi} = 8 c_{\rm s}$) in Regions II and III in Figure
\ref{fig:unstable_regions_num} are clearly identified.  The complete
mode structure corresponding to this case can be seen 
in the left panels of Figure \ref{fig:approx}.

It is important to stress that neither the real nor the imaginary parts of
either $R_1$ or $R_2$  are negligible compared to unity even for
\Alfven speeds of order a few times the sound speed.  In fact, for the
unstable modes, the ratio $R_1$ is of order unity   and the ratio
$R_2$ is in some cases larger than one by one order of magnitude. Their
functional form is significantly different than the assumed $1/k_z$.

\subsection{Magnetorotational Instabilities with Superthermal Fields}
\label{subsec:mri_superthermal}

\begin{figure}[tbh]
\begin{center}
\includegraphics[width=.475\textwidth]{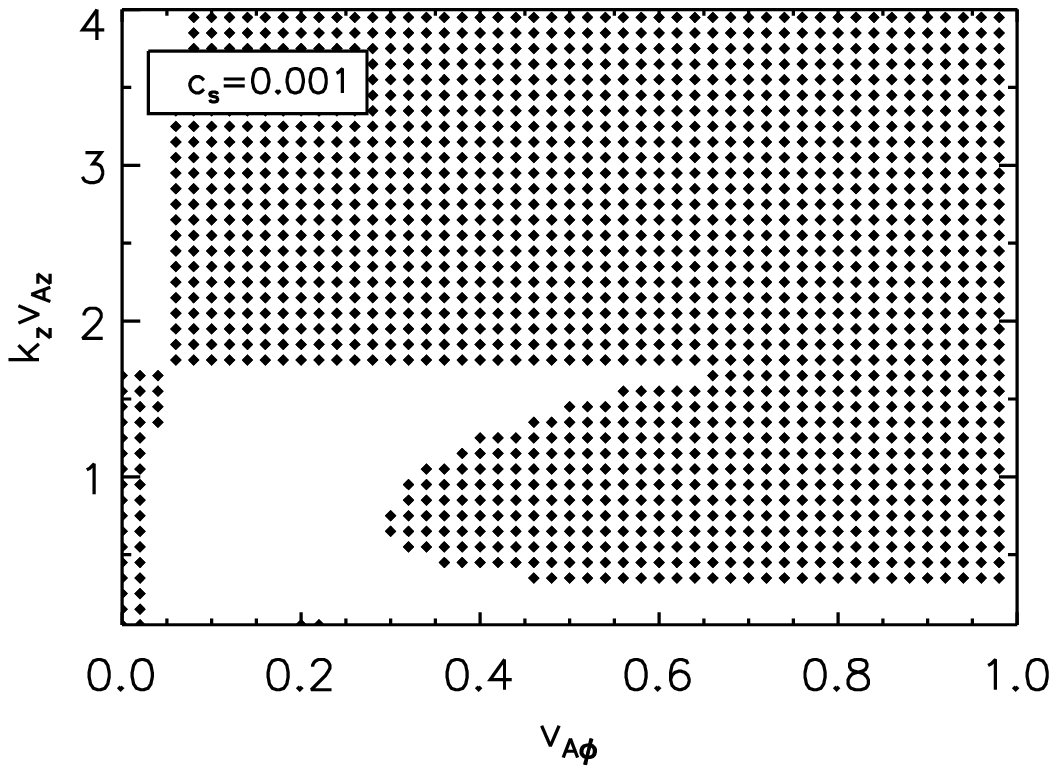}
\includegraphics[width=.475\textwidth]{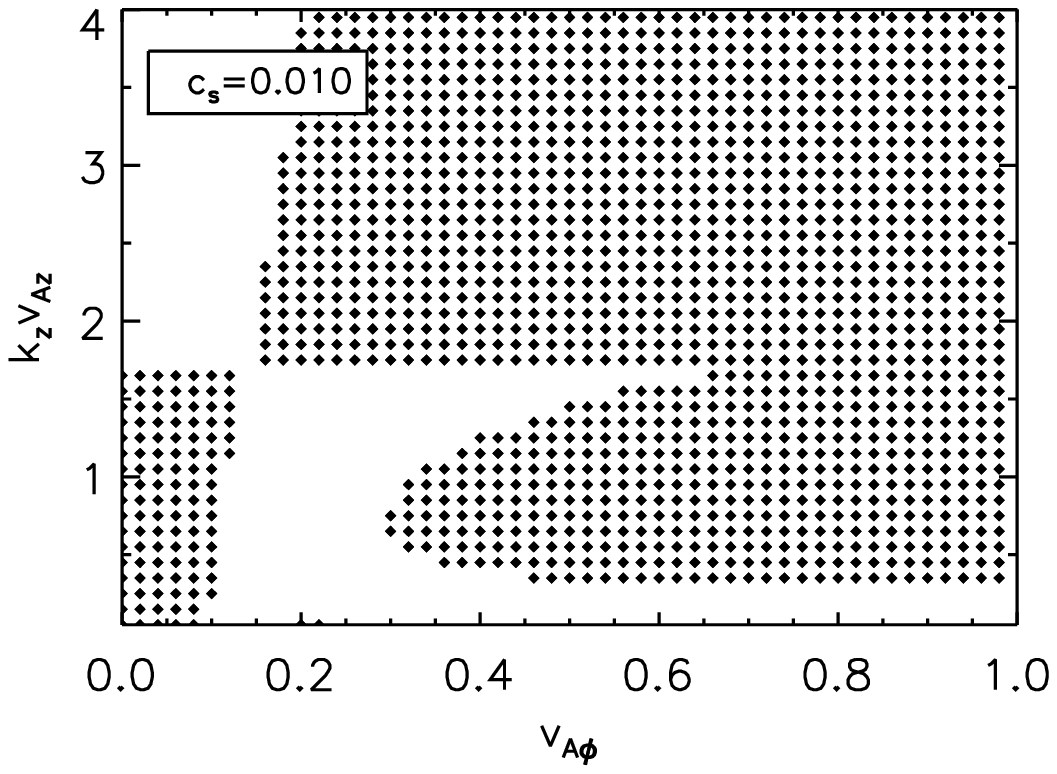}
\includegraphics[width=.475\textwidth]{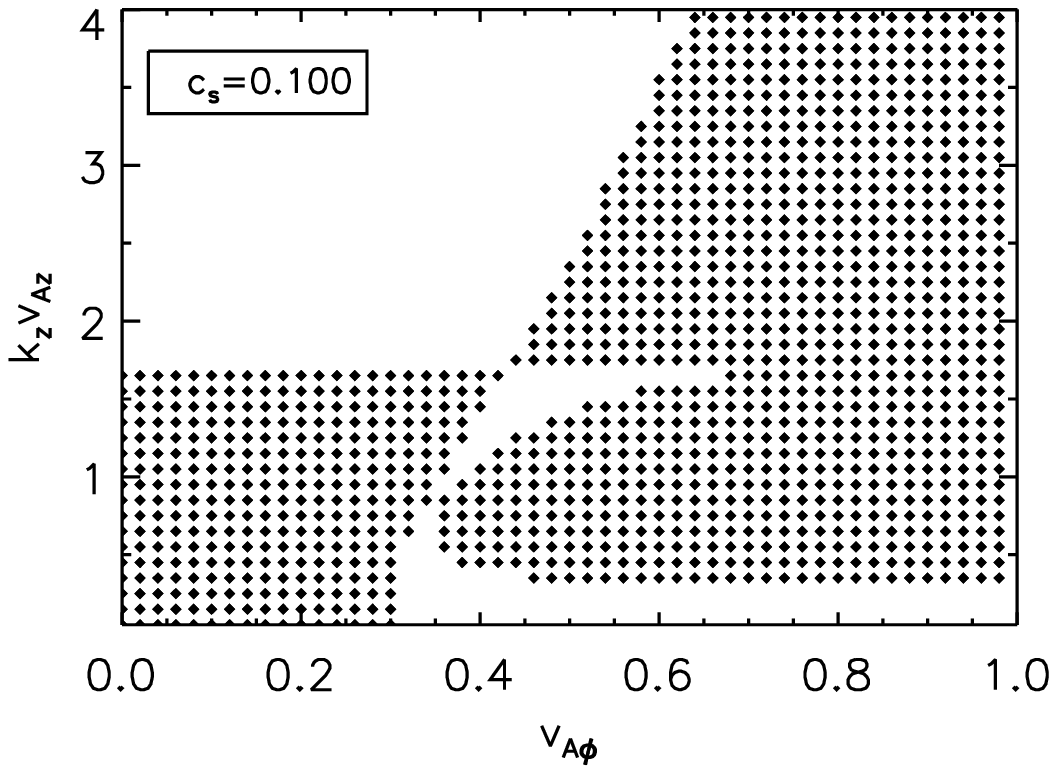}
\includegraphics[width=.475\textwidth]{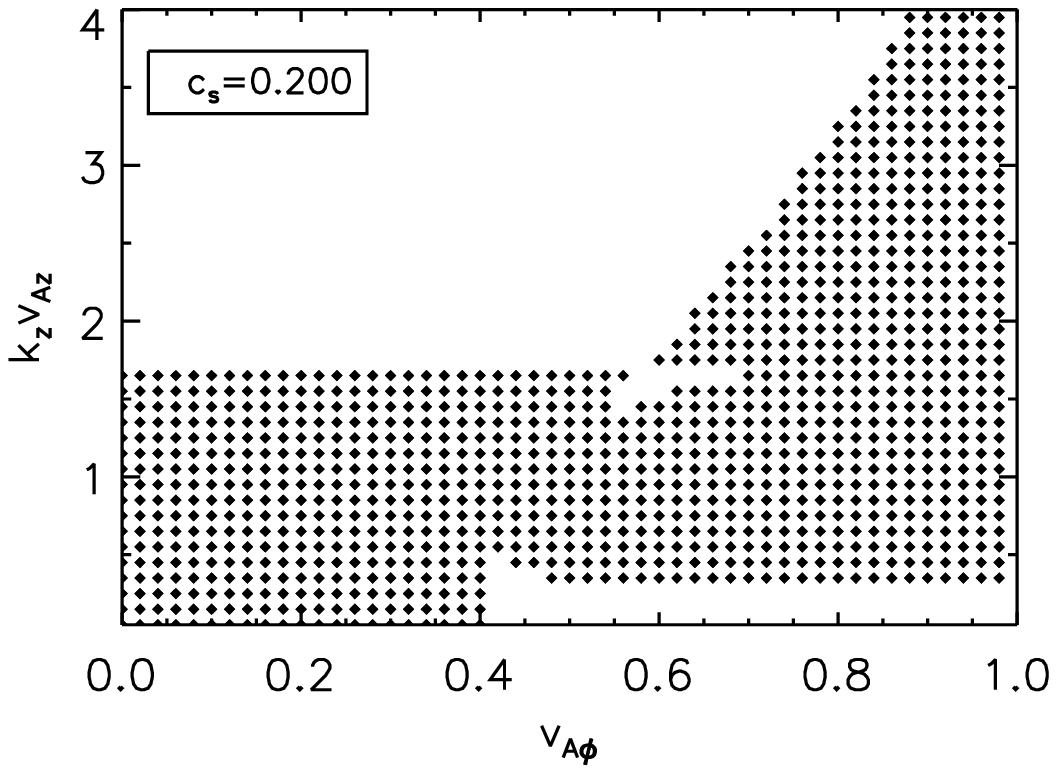}
\caption{The black dots represent unstable modes obtained from solving the
        dispersion relation (\ref{eq:disp_full_nodim_kr}) numerically, for
        a Keplerian disk with $v_{{\rm A}z}= 0.01$. In each plot, 
	different values of the local sound speed, $c_{\rm s}$, are
        considered. Note that in this case, the axes are not normalized
        by the particular value of the local sound speed, but rather
        by our initial choice of dimensionless variables,
	see \S \ref{sec:mhd equations and dispersion relation}.}
\label{fig:sound}
\end{center}
\end{figure}

In \S\ref{sec:numerical solutions} we demonstrated that, when the
toroidal magnetic field in a differentially rotating MHD flow becomes
superthermal, three distinct instabilities can be identified, which we
denote by roman numerals I, II, and III in
Figure~\ref{fig:unstable_regions_num}. We summarize the 
qualitative characteristics of these instabilities below.

In contrast to the weak-field MRI, all three instabilities correspond
to compressible MHD modes. Moreover, while the traditional MRI
corresponds to perturbations with negligible displacements along the
vertical direction, this is not true for any of the three
instabilities with superthermal toroidal fields. Instead, vertical
displacements are an important characteristic of these instabilities
and they occur with negligible acceleration, under a force balance
between thermal and magnetic pressure. Finally, as in the case of the
MRI, there is no significant acceleration along the radial direction
but rather a force balance between magnetic tension, magnetic
pressure, and thermal pressure.

In Figures~\ref{fig:dlog} and \ref{fig:sound} we study numerically the
dependences of the three instabilities on the properties of the
background flow. As also shown in the case of the weak-field MRI
(Balbus \& Hawley, 1991),  instability I occurs only in 
differentially rotating flows, with radially decreasing angular velocity. 
However, instability I also requires the presence of a
non-negligible thermal pressure. Either a radially increasing angular
velocity or a superthermal toroidal field can suppress instability I
and hence the traditional MRI. 

Instability II is ubiquitous, whenever the background toroidal field
of the flow is significantly superthermal. Indeed, it occurs even for
flat (see Fig.~\ref{fig:dlog}) or very cold (see Fig.~\ref{fig:sound})
flows. In a sense opposite to instability I, the steepness of the
rotational profile determines the minimum unstable wavenumber, whereas
the magnitude of the sound speed determines the minimum toroidal field
strength required for the instability to occur.
This instability seems to correspond to a generalization of the
axisymmetric toroidal buoyancy (ATB) modes identified in \cite{KO00},
where the case $c_{\rm s}=0$ was studied.  In a similar way to
instability II, the ATB modes with $c_{\rm s}=0$ become unstable
for all wavenumbers exceeding a critical value (for vertical modes this
value is just given by $k_{\rm BH}$).  
When a finite sound speed is considered, however, 
thermal effects play an important
role at small scales by completely stabilizing all the modes with
wavenumbers larger than $k_z^{{\rm c} 2}$ (eq.~[\ref{eq:kc2}]). 

Finally, instability III depends strongly on the rotational profile
but very weakly on the sound speed. For rotationally supported flows
(i.e., for $v_{\rm A \phi}\ll 1$), instability III occurs only for
significantly steep rotational profiles, e.g.,  
$q=|d\ln \Omega/ d\ln r|\gtrsim 1.0$, for
the parameters depicted in Figure~\ref{fig:dlog}.

\newpage

\begin{figure}[tbh]
\begin{center}
\includegraphics[width=\textwidth]{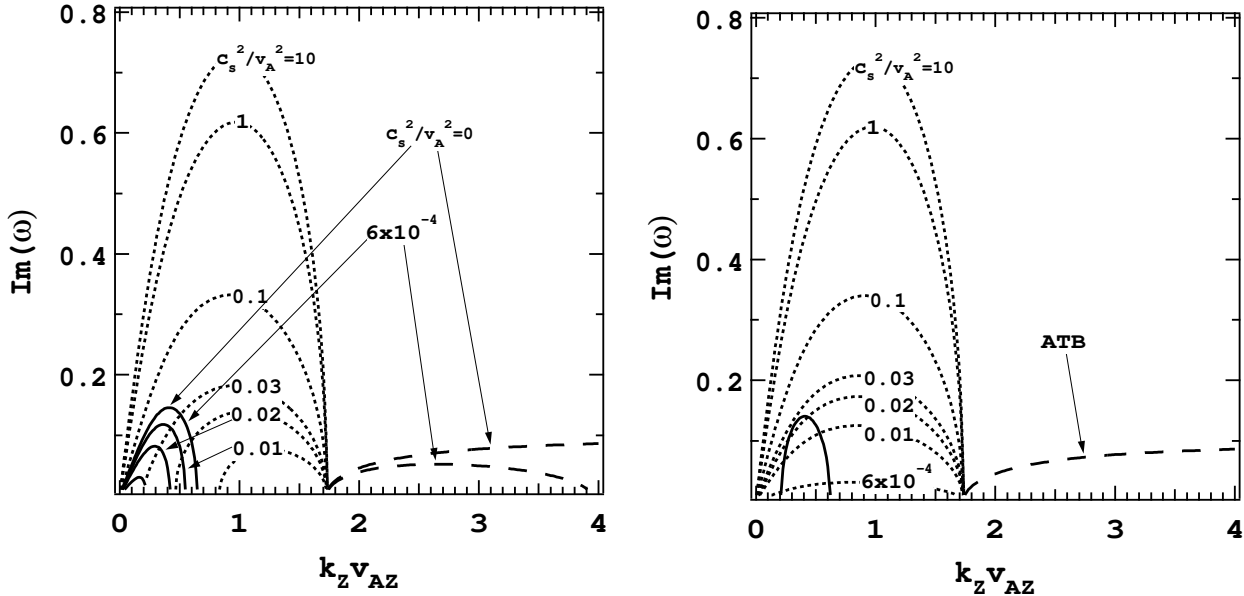}
\caption{The growth rate evolution of the different instabilities
defined in \S \ref{subsec:unstable modes} for increasing magnetic
field strength parameterized in terms of  the ratio $c_{\rm s}^2/
v_{{\rm A}}^2$ for a fixed pitch angle  $i\equiv\tan^{-1} (v_{{\rm A}
z}/v_{{\rm A} \phi}) =25 \degr$.  \emph{Left panels}: Growth rates of
the unstable solutions to the full dispersion relation
(\ref{eq:disp_full_nodim_kr}), when all curvature terms are taken into
account. \emph{Right panels}: Growth rates of the unstable solutions
to the dispersion relation (\ref{eq:disp_cseq0}),  i.e., when all
curvature terms are neglected. In both cases, we have considered a
vanishing ratio $k_r/k_z$, $v_{{\rm A} z}=0.05$, and a  Keplerian
disk.  The dotted lines, in both panels,  show the 
stabilization of the standard MRI 
as the magnetic field becomes superthermal (instability
I). The solid and dashed lines on the left panel show the growth rates
corresponding to instability III and II respectively. As discussed in
\S \ref{sec:numerical solutions}, these instabilities do not have a
counterpart when the magnetic tension induced by bending of toroidal
field lines is neglected.  For completeness we have included, in the
right panel, with solid and dashed lines the unstable solutions to
equation (44) in Kim \& Ostriker (2000). Instability II corresponds to
a generalization of the ATB mode when thermal effects are accounted for.}
\label{fig:growth_rates}
\end{center}
\end{figure}

\subsection{Comparison to Previous Analytical Studies}
\label{subsec:Comparison to previous analytical studies}

Soon after the original paper by \citet{BH91}, \citet{Knobloch92}
critiqued their approach to the study of local instabilities
for lacking the contributions of curvature terms.
\citet{Knobloch92} 
formulated the stability analysis of a vertically unstratified,
incompressible disk as an eigenvalue problem in the radial
coordinate. He found that the presence of a toroidal field
component changes the conditions for the presence of the instability
as well as the character of the unstable modes from purely exponentials to
overstable (i.e., $Re[\omega] \ne 0$). \citet{GB94} argued against
Knobloch's findings regarding overstability, stating that it arose as a
consequence of having kept only  small order terms (like
$v_{\rm A}/c_{\rm s}$ and $v_{\rm A}/\Omega_0 r_0$).  They concluded
that these contributions would have been negligible had the flow been
considered compressible.

As we comment in \S \ref{subsec:analytic approximations},  Knobloch's
dispersion relation is correct even in the limit $c_{\rm s} \gg 
v_{{\rm A}\phi}$  (i.e., without the necessity of imposing strict
incompressibility).  Formally speaking, the linear term in $\omega$ in
equation (\ref{eq:disp_incomp_nodim}) does break the symmetry of the
problem allowing for unstable modes with $Re(\omega) \ne 0$.  But it
is also the case that, in the limit  $c_{\rm s} \gg v_{{\rm A}\phi}$, because
of the relative magnitude of the coefficients in the dispersion
relation (\ref{eq:disp_incomp_nodim}), we do not expect the stability
properties of the flow to differ greatly from those described by the
incompressible MRI.  As we mention in \S \ref{sec:previous treatments},
in order to see significant differences, the \Alfven speed would 
have to be of the
order of the circular speed and therefore we do not expect the
curvature terms to play a significant role on the stability of
incompressible, rotationally supported flows.
On the other hand, if we allow the MHD fluid to be compressible and
consider the curvature of the background flow, the
mode structure can be radically different from what is expected for
the compressible MRI (c.f. Blaes \& Balbus 1994). This is the case, even if the
toroidal \Alfven speed exceeds the sound speed by a factor of a few
without the necessity of violating the condition of a rotationally
supported disk (see Fig. \ref{fig:comp_real}). 

The stability of axisymmetric perturbations in weakly ionized and
weakly magnetized shear flows was considered by \citet{BB94}. They
showed that, when ionization equilibrium is considered in the
two-fluid approach, strong toroidal fields can fully stabilize the
flow. As part of their study, they relaxed the Boussinesq
approximation in the case of a single fluid and argued that, to all
orders in the field strength, the magnitude of  
$B_\phi$ does not affect the stability
criterion.  
As noted by \citet{CP95}, this conclusion was reached
because the terms proportional to
$B_\phi/r_0$ were not included in the local analysis.

The behavior of the MRI in cold MHD shearing flows,  has been
addressed by Kim \& Ostriker (2000).  When performing their local
analysis, these authors obtained the compressible version of the standard
dispersion relation for the MRI and studied its solutions  for
different values of the ratio $c_{\rm s}^2/v_{\rm A}^2$.  
Analyzing their dispersion relation (i.e.,
their equation [57] which is equivalent to equation [\ref{eq:disp_cseq0}] in
this study), Kim \& Ostriker (2000) concluded that, when the magnetic
field is superthermal, the inclusion of a toroidal component
suppresses the growth rate of the MRI. Moreover, they found that, 
for a Keplerian rotation law, 
no axisymmetric MRI takes place in very cold MHD flows
if $i<30\degr$, where $i$ is the pitch angle of the local 
magnetic fields, $i\equiv\tan^{-1}(v_{{\rm A} z}/v_{{\rm A} \phi})$.

\begin{figure}[tbh]
\begin{center}
\includegraphics[width=\textwidth]{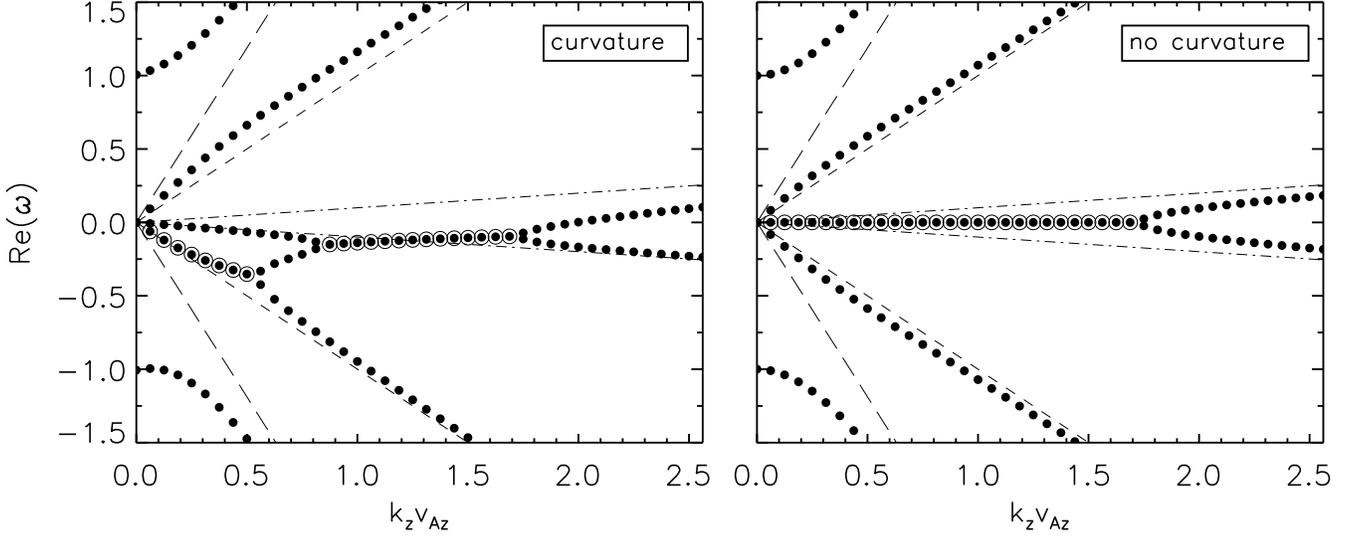}
\caption{The real parts of the solutions to the dispersion relation
(\ref{eq:disp_full_nodim_kr})  corresponding to  $c_{\rm
s}^2/v_{{\rm A}}^2=0.01$  for a pitch angle   $i=\tan^{-1} (v_{{\rm A}
z}/v_{{\rm A} \phi}) =25 \degr$, a vanishing ratio $k_r/k_z$,
$v_{{\rm A} z}=0.05$, and a Keplerian disk.  \emph{Left
panel}: Solutions to the full dispersion relation
(\ref{eq:disp_full_nodim_kr}), when all curvature terms are taken into
account. \emph{Right panel}: Solutions  to the dispersion relation
(\ref{eq:disp_blaes}), i.e., when all curvature terms are
neglected. Open circles indicate unstable modes.  In
both cases, the unstable modes with the shortest  wavelength correspond to
$k_{\rm BH} v_{{\rm A} z}$.}  
\label{fig:mri_stable}
\end{center}
\end{figure}

In  \S \ref{sec:onset of instabilities} we showed that,
depending on the strength of the toroidal field component,
accounting for the finite curvature of the background magnetic field
and the finite compressibility of the flow could be crucial in
establishing which modes are subject to instabilities.
In particular,
we stated that both effects should be considered simultaneously
whenever the local value of the toroidal \Alfven speed exceeds the
geometric mean of the local sound speed and the local rotational
speed (for a Keplerian disk). 
However, this analytic criterion was found to be relevant for
the modes with frequencies satisfying the condition $\omega^2 \ll
k_z^2 c_{\rm s}^2$.   Therefore, it is not obvious that we can trust
this criterion in the  limit $c_{\rm s} \rightarrow 0$.

In order to
see whether finite  curvature effects do play a role in the stability
of cold MHD  flows we solved the complete  dispersion relation
(\ref{eq:disp_full_nodim_kr}) for a pitch angle $i=25\degr$ and
considered different values for the ratio $c_{\rm s}^2/v_{\rm
A}^2$. 
Figure \ref{fig:growth_rates} shows the growth rates for the
unstable modes of our dispersion relation (i.e.,
eq.~[\ref{eq:disp_full_nodim_kr}]) and  compares them to the ones of
the dispersion relation obtained when the curvature terms are
neglected (i.e., eq.~[\ref{eq:disp_cseq0}]).
In both cases, we have considered
$k_z \gg k_r$, $v_{{\rm A} z}=0.05$, and a  Keplerian disk. The
stabilization of the standard MRI (i.e., instability I) as the
magnetic field becomes superthermal ($v_{\rm A}>c_{\rm s}$)  is
evident (dotted lines in both panels).  When the effects of magnetic
tension are considered, not only does the growth rate of the MRI
decrease faster for low values of  $c_{\rm s}^2/v_{\rm A}^2$ but the
modes with longest wavelengths are no longer unstable
(e.g. when   $c_{\rm s}^2/v_{\rm A}^2=0.01$). Because of
this, the  MRI is completely stabilized even for finite values of
$c_{\rm s}^2/v_{\rm A}^2$. In contrast, when the curvature of the 
field lines is neglected, the growth rates decrease but the range
of unstable modes remains unchanged as  $c_{\rm s} \rightarrow 0$
(right panel in Figure \ref{fig:growth_rates});  
it is only when $c_{\rm s} = 0$ that the MRI is completely suppressed. 
For completeness, we present in Figure \ref{fig:mri_stable}  the real parts
of the  solutions to the dispersion relation
(\ref{eq:disp_full_nodim_kr})  for $i=25\degr$, $k_z \gg k_r$,  $v_{{\rm
A} z}=0.05$, $c_{\rm s}^2/v_{\rm A}^2=0.01$ and a Keplerian disk, in
the cases where curvature terms are considered (left panel) and
neglected (right panel). The stabilization of the MRI at low
wavenumbers and the emergence of instability III are evident.
Note that, from Figure \ref{fig:mri_stable}, 
it is clear that the inclusion of magnetic tension 
terms can cause modifications to the mode structure
when $v_{\rm A} \gg c_{\rm s}$, 
even  when the toroidal and vertical components of the
magnetic field are comparable.

In \S \ref{subsec:mri_superthermal} we mentioned that instability
II seems to be a generalization of the axisymmetric toroidal buoyancy
(ATB) modes, identified by Kim \& Ostriker (2000), that accounts for
finite temperature effects. Further indication that this is indeed the
case can be found in Figure \ref{fig:growth_rates} where we have
plotted (dashed lines) the growth rates corresponding to instability
II and the one corresponding to the ATB mode (solutions  of equation
[44] with $\omega^2 \ll k_z^2 v_{{\rm A} z}^2$  in  Kim \& Ostriker
2000).  Although finite compressibility suppresses instability II at
large wavenumbers, it is clear that, as $c_{\rm s} \rightarrow 0$, the
growth rates associated with instability II tend continuously to the
growth rate of the cold ATB mode.  For completeness, we have also
included, in the right panel of Figure \ref{fig:growth_rates},  the
growth rate corresponding to the remaining unstable solution  of
equation (44) in  Kim \& Ostriker (2000) (solid line).  This growth
rate should be interpreted with great care since the aforementioned
equation was derived
under the condition $\omega^2\ll k_z^2 v_{\rm A}^2$, which is not
satisfied by the corresponding unstable mode.   Nonetheless, we have
included it to show the similarities that it shares with the growth
rates corresponding to instability III (solid lines in the left panel)
as  $c_{\rm s} \rightarrow 0$ .  Note that the growth rates
corresponding to instability III increase as  $c_{\rm s} \rightarrow
0$ and they saturate at  $c_{\rm s} =0$.  Although the higher critical
wavenumber and the growth rate around this critical wavenumber seem to
be the same for both instabilities, the differences between them at
low wavenumbers is also evident. These differences become more
dramatic as the pitch angle increases.

Curvature terms cannot, of course, be neglected in global treatments
of magnetized accretion disks.  It is, therefore, not surprising that
new  instabilities, distinct from the MRI, have already been found in
global studies in which strong fields were considered.  In
particular, \citet{CP95} performed a global stability analysis to
linear axisymmetric perturbations of an incompressible, differentially
rotating fluid, threaded by vertical and toroidal fields. They
considered power-law  radial profiles for the angular velocity and the
toroidal and vertical components of the field. Each of these were
parameterized as  $\Omega \propto r^{-a}$, $B_\phi \propto r^{-b+1}$,
and $B_z \propto r^{-c+1}$, respectively. Most of their analysis dealt
with a constant vertical field and they allowed variations of the
exponents $(a,b)$,  with the restriction that they correspond to a
physical equilibrium state with a stationary pressure distribution.
Although the majority of that paper dealt with global characteristics,
they also performed a WKB analysis and concluded that, for $ 3/2 \le
a=b \le 2$  and $v_{{\rm A} \phi} < 1 $, the growth rate of unstable
modes is suppressed on both short and long wavelengths and it
approaches zero when $v_{{\rm A} \phi} \rightarrow 1$. On the other
hand, for  $ a = b \ne 2$ and $v_{{\rm A} \phi} > 1 $,  they found a
new instability, with a growth rate that increased with $v_{{\rm A}
\phi}$. They call this the Large Field Instability (LFI) and showed
that it can be stabilized for sufficiently large  $v_{{\rm A} z}$.

It is worth mentioning the major qualitative differences between the
LFI and the new instability discussed in \S \ref{sec:numerical
solutions} that emerges for $k_z < k_{\rm BH}$ after the stabilization
of the MRI. Although it is true that, for our instability to be
present, it is necessary for the toroidal \Alfven speed to exceed the
local sound speed, there is no need to invoke \Alfven speeds larger
than the local rotational speed. This is in sharp contrast with the
LFI which only appears for $v_{{\rm A} \phi} > 1$. Regarding the range
of unstable wavenumbers,  the LFI remains unstable for $k_z \rightarrow
0$, albeit with diminishing growth rate for large values of $v_{{\rm
A} z}$. This is not the case for the new instability present at low
wavenumbers in our study.  This can be seen, for example, in the
left panels in Figure \ref{fig:approx}. Perhaps
the most noticeable difference is that the two instabilities in
\citet{CP95} that are present in the case $a = b \ne 2$ do not seem
to coexist under any particular circumstances. The instability present
for $v_{{\rm A} \phi} < 1$ reaches zero growth for $v_{{\rm A} \phi}
\rightarrow 1$, while the LFI appears for $v_{{\rm A} \phi} =1$ and
its growth rate is proportional to $v_{{\rm A} \phi}$. When
compressibility is considered, however, the two new instabilities found in
our study can coexist even for \Alfven speeds smaller than the local
rotational speed.

\subsection{Implications for Shearing Box Simulations}
\label{sec:implications}

In an attempt to capture the most relevant physics without all the
complexities involved in global simulations,  the shearing box
approach has been widely used in numerical studies of magnetized
accretion disks (see, e.g., Hawley, Gammie, \& Balbus 1994).  The aim
of the shearing box approximation is to mimic a small region of a
larger disk. The size of the box is usually  $H_z \times 2\pi H_z
\times H_z$, with $H_z$ the thermal scale height of the isothermal
disk.  In this approach, it is common to adopt a pseudo-Cartesian
local system centered at  $r_0$ and in corotation with the disk with
an angular frequency $\Omega_0$, with coordinates $x=r-r_0$,
$y=r_0(\phi-\Omega_0 t)$, and $z$. The effects of differential
rotation are then considered by imposing a velocity gradient in the
radial direction. For a Keplerian accretion disk this is achieved by
setting $v_y= -(3/2) \Omega_0 x$.

\begin{figure}[tbh]
\begin{center}
\includegraphics[width=0.7\textwidth, angle=90]{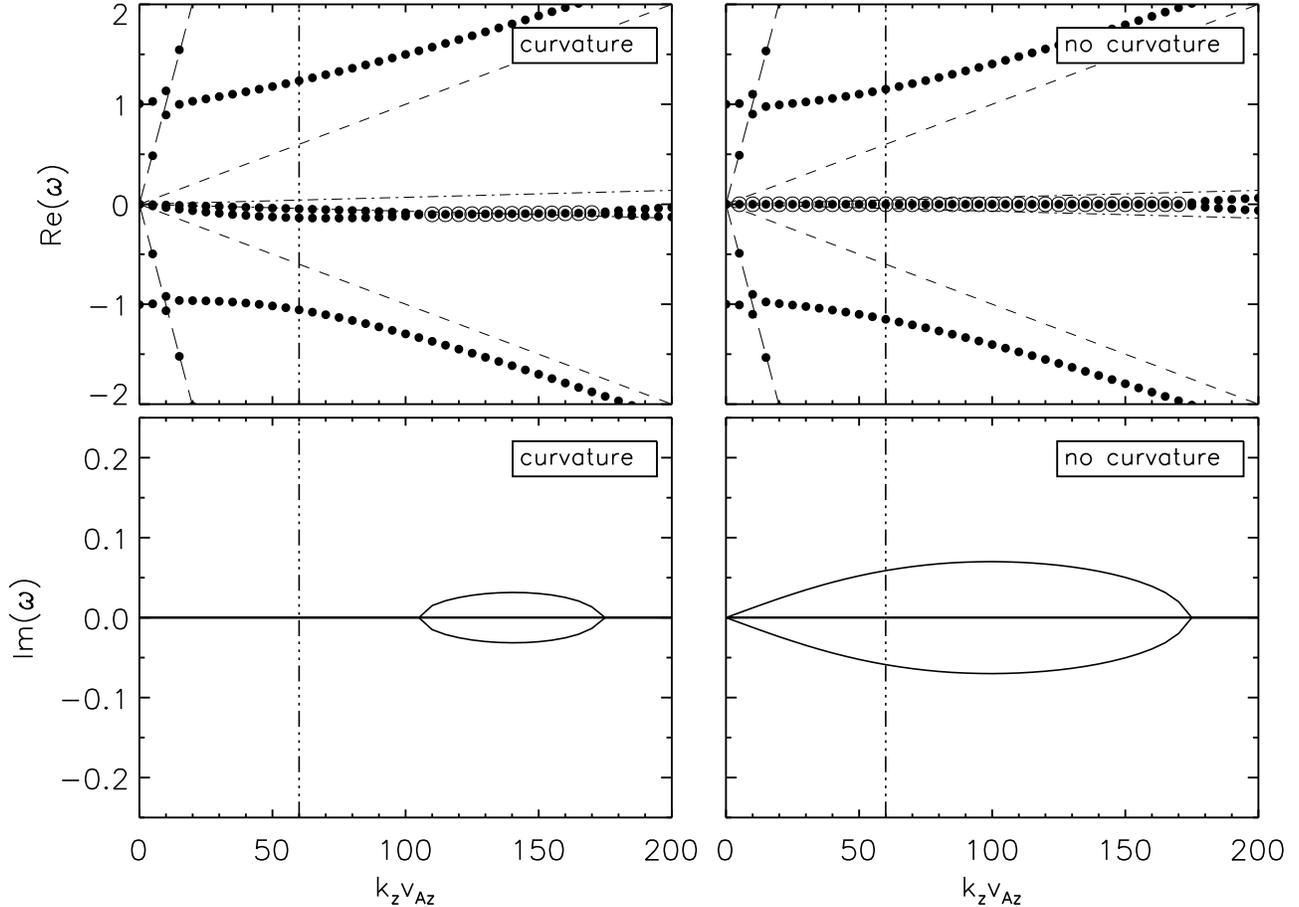}
\caption{The implication of our study for shearing box simulations.
\emph{Left panels}: Solutions to the full dispersion relation
(\ref{eq:disp_full_nodim_kr}), when all curvature terms are taken into
account. \emph{Right panels}: Solutions to the dispersion relation
(\ref{eq:disp_blaes}), i.e., when all curvature terms are
neglected. In both cases, we have considered 
$c_{\rm s}= 0.007$, $v_{{\rm A}z}=0.01$, $v_{{\rm A} \phi}=0.1$, 
and a Keplerian disk.
Open circles in upper panels indicate unstable modes. The vertical
line indicates the minimum wavenumber (i.e., largest wavelength) that
can be accommodated in the simulations of a strongly-magnetized corona
above a weakly magnetized disk by \citet{MS00}.}
\label{fig:shearing_box}
\end{center}
\end{figure}

In most studies of unstratified shearing boxes, \Alfven speeds rarely
exceed the value of the local sound speed (see, e.g., Hawley, Gammie,
\& Balbus 1995, 1996). This is mainly because they are designed to
simulate the mid-plane of the disk where the flow is relatively
dense. In \S \ref{subsec:analytic approximations}, we have seen that,
as long as the toroidal \Alfven speed does not exceed the critical
value $v_{{\rm A} \phi}^2 = c_s \kappa $, neglecting  magnetic tension
due to the curvature of toroidal field lines does not seem to affect the
outcome of the MRI and hence the shearing box approach is well
justified.  However, when stratification is taken into account,
usually by adopting a density profile of the form $\rho \propto
\exp[-z^2/(2H^2)]$ in the case of isothermal disks, the steep drop in
the density beyond a few scale heights can potentially lead to a
magnetically dominated flow, with \Alfven speeds larger than the
critical value $v_{{\rm A} \phi}^2 = \kappa c_{\rm s}$.

As discussed in the introduction, \citet{MS00} carried out
three-dimensional MHD simulations to study the evolution of a
vertically stratified, isothermal, compressible, magnetized
shear flow.  The simulations were local in the plane of the disk but
vertically extended up to $\pm 5$ thermal scale heights. This allowed
them to follow the highly coupled dynamics of the weakly magnetized
disk core and the rarefied magnetically-dominated (i.e., $\beta < 1$)
corona that formed above the disk, for several (10 to 50) orbital
periods.

Miller \& Stone (2000) considered a variety of models, all sharing the same initial
physical background, but different initial field configurations.  In
particular, they considered the following values: $\Omega_0=10^{-3}$,
$2c^2_{\rm s}=10^{-6}$ (so that $H_z=\sqrt{2} c_{\rm s}/\Omega_0= 1$),
and $r_0=100$. We mention some of the results they obtained for the
models with initial toroidal fields (BY), which were qualitatively
similar to the zero net $z$-field (ZNZ) models. After a few orbital
periods, the presence of a highly magnetized (with plasma $\beta
\simeq 0.1 - 0.01$) and rarefied (with densities two orders of
magnitude lower than the disk mid-plane density) corona above $\sim 2$
scale heights is evident.  Within both the disk and the corona, the
``toroidal'' component of the field ($B_y$), favored by differential
rotation, dominates the poloidal component of the field by more than
one order of magnitude (with $B_x^2 \simeq B_z^2$).

We can compare the predictions of our study to the mode structure that
one might expect from the standard  compressible MRI for the
particular values of sound and \Alfven speeds found in the strongly
magnetized corona by \citet{MS00}.  To this end, we consider as
typical (dimensionless) values  $c_{\rm s}=0.007$, $v_{{\rm A}
\phi}=0.1$ and $v_{{\rm A} z}=0.01$, where we have assumed $\beta =
2(c_{\rm s}/ v_{\rm A})^2 \simeq 0.01$ and $B_\phi=10 B_z$.   
The largest features that their simulations are able
to accommodate are those with $k_z \sim 60$ (corresponding to a
wavelength of 10 in the vertical direction). As it can be seen in
Figure \ref{fig:shearing_box},  the role of the curvature terms is not
negligible in two different respects. First, it completely stabilizes
the perturbations on the longest scales well inside the numerical
domain. Second, the growth rate for the unstable modes is
significantly reduced.

It is difficult to extrapolate from the present work to address how
the instabilities discussed here would couple to buoyancy in the
presence of a stratified medium like the one considered by
\citet{MS00}. Shearing boxes might also suffer from other problems
when used to model strongly magnetized plasmas (e.g., the shearing
sheet boundary conditions in the radial direction might not be
appropriate for strong fields). The question is raised, however, about
whether, because of their own Cartesian nature, they constitute a good
approach at all to simulating compressible flows in which superthermal
toroidal fields are present.  Despite the fact that the generation of
strongly magnetized regions via the MRI in stratified disks seems hard
to avoid, their stability properties will ultimately depend on both
the use of proper boundary conditions and  proper accounting of the
field geometry.

\section{\textsc{SUMMARY AND CONCLUSIONS}}
\label{sec:summary and conclusions}

In this paper we have addressed the role of toroidal fields on the
stability of local axisymmetric perturbations in compressible,
differentially rotating, MHD flows, when the geometrical curvature  of
the background is taken into account. In order to accomplish this
task, without imposing restrictions on the strength of the background
equilibrium field, we relaxed the Boussinesq approximation.  In
particular, we have studied under which circumstances the curvature
terms, intimately linked to magnetic tension in cylindrical coordinate
systems, cannot be neglected.  We have shown that the MRI is
stabilized and two distinct instabilities appear for strong toroidal
fields.  At least for large wavenumbers,  the structure of the modes
seems to be the result of a purely local effect that is accounted for
when compressibility and curvature terms are consistently taken into
account. In particular, we have demonstrated that, even for
rotationally supported cylindrical flows, 
both curvature terms and flow compressibility  have to be
considered if, locally,  the toroidal \Alfven speed exceeds 
the critical value given 
by $v_{{\rm A} \phi}^2 = (\kappa/\Omega) c_{\rm s} \Omega r $ 
(in physical units).

There is little doubt that a realistic treatment of normal modes in
magnetized accretion disks has to contemplate gradients in the flow
variables over large scales and should, therefore, be global in
nature.  The results presented in this paper, however,  provide the
complete dispersion relation and, more importantly, analytic
expressions for some of its solutions that should be recovered, in
the appropriate limit, by a study of global modes in  magnetized
accretion disks, where compressibility effects are likely to be non
negligible. We will address these issues in a future paper.

\begin{figure}[tb]
\begin{center}
\includegraphics[width=\textwidth]{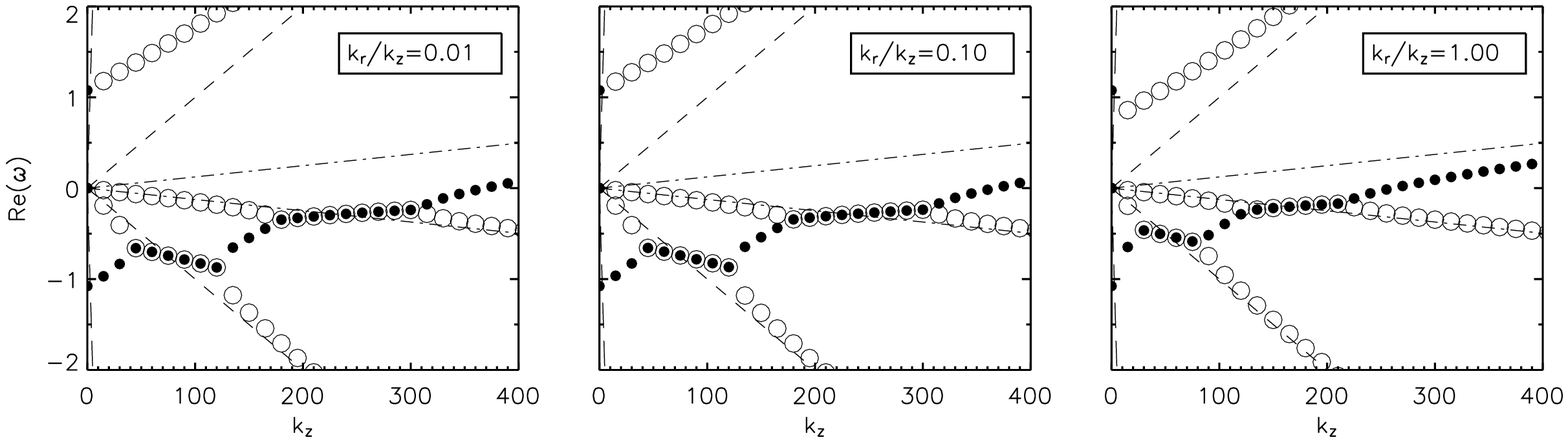}
\includegraphics[width=\textwidth]{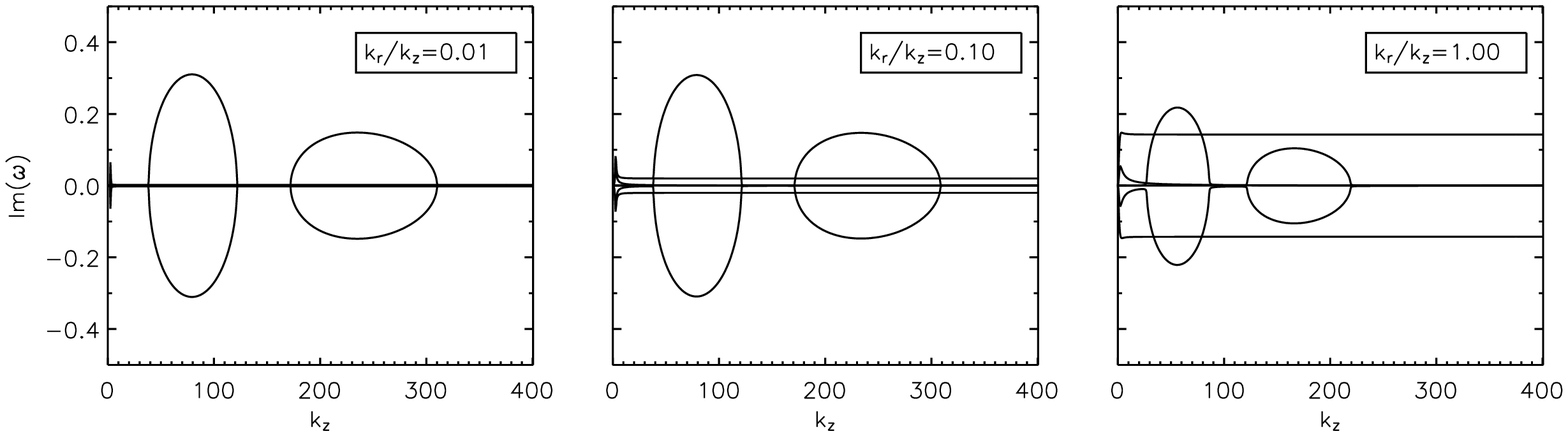}
\caption{Solutions to the dispersion relation
(\ref{eq:disp_full_nodim_kr})  for different values of the ratio
$k_r/k_z$. The left, central, and right panels show the results for
$k_r/k_z=0.01, 0.1$, and 1 respectively.  In all three cases, we have
considered $c_{\rm s}=0.05$, $v_{{\rm A} z}=0.01$, $v_{{\rm A}
\phi}=0.4$, and a Keplerian disk.  Open circles
in upper panels indicate unstable modes.  The same instabilities that
were present in the case  with vanishing ratio $k_r/k_z$ 
(left panels in Figure  \ref{fig:approx}) can be clearly identified.}

\label{fig:kr_ne_0}
\end{center}
\end{figure}

\newpage

\section*{\textsc{APPENDIX A}}
\label{sec:APPENDIX A}

The central dispersion relation obtained in this study,  i.e.,
equation (\ref{eq:disp_full_nodim_kr}), was derived considering
axisymmetric perturbations in both the vertical and the radial
directions. Throughout the majority of our analysis,  however, we
focused our attention  on the modes with vanishingly small ratios
$k_r/k_z$. Particular emphasis has been given to the study of these
modes in the literature of weakly magnetized, differentially rotating
disks, since these are the modes that exhibit the fastest growth rates
\citep{BH92, BH98, Balbus03}.   After having analyzed the role played
by magnetic tension forces due  to the finite curvature of strong
toroidal field lines on the stability of these modes, we are in a
better position to understand their effects on the modes for which the
ratio $k_r/k_z$ is finite.

In Figure \ref{fig:kr_ne_0}, we present the solutions to the dispersion
relation (\ref{eq:disp_full_nodim_kr}) for three different ratios of
the radial to the vertical wavenumber, i.e., $k_r/k_z = 0.01, 0.1$,
and 1. For the sake of comparison, we have used in this figure the
same parameters that we used in obtaining  the left panels in Figure
\ref{fig:approx}, for which the ratio $k_r/k_z$ was considered to be
vanishingly small;
we have assumed $c_{\rm s}= 0.05$, $v_{{\rm A}z}=0.01$,
$v_{{\rm A} \phi}=0.4$ and a Keplerian disk.   In all the cases, the
same instabilities (II and III) that were present in the case $k_z \gg
k_r$ can be clearly identified.   Note however, that even for very
small ratios $k_r/k_z$ (e.g., left panels in Figure \ref{fig:kr_ne_0})
some of the modes that were stable in the case $k_z \gg k_r $ 
become unstable,  albeit with negligible growth rate,
when compared with the other unstable modes.  It is also evident that
in the limit $k_z \gg k_r$, the mode structure in Figure
\ref{fig:kr_ne_0} tends continuously toward the mode structure in the
left panels in Figure \ref{fig:approx}. It is this continuous behavior
that ultimately  justifies the study of modes with negligible ratio
$k_r/k_z$ in a local stability analysis.

In the case $k_r =k_z$ (right panels in Figure \ref{fig:kr_ne_0}),
the value of the critical vertical wavenumbers
for the onset of instabilities, i.e., $k_z^{{\rm c} 1}$ and $k_z^{{\rm c}
2}$ in  equations (\ref{eq:kc1}) and (\ref{eq:kc2}) respectively, are
different with respect to the case in which $k_z \gg k_r$ by a factor
$\sqrt{2}$. This  indicates that $k_z$ and $k_r$ play similar roles in
establishing these critical wavenumbers. The growth rates of all these
modes are reduced with respect to the case with  $k_z \gg k_r$.
This behavior is similar to the one observed in the case of weak
magnetic fields.  It is important to stress that,  even for the modes
with comparable values of vertical and radial wavenumbers, 
the general characteristics of
the  instabilities for strong toroidal fields  that we discussed in \S
\ref{subsec:unstable modes}  are insensitive to the inclusion of a
non-negligible $k_r$.

The completely new feature in Figure \ref{fig:kr_ne_0}  is the
appearance of another instability with a growth rate that does not
seem to depended on wavenumber; the terms  proportional to $ik_r$ in
equation (\ref{eq:disp_full_nodim_kr}) are crucial for the appearance
of this new instability.  The mode that is unstable seems to
correspond to the mode that becomes the \Alfven mode in the limit of
no rotation.  With increasing values of the ratio $k_r/k_z$, the
growth rates of the instabilities studied in \S \ref{subsec:unstable
modes} go to zero, but the new instability in Figure \ref{fig:kr_ne_0}
persists.  Note that for $k_r \simeq k_z$ the growth rates of all the
instabilities in  Figure \ref{fig:kr_ne_0} are comparable.

As an aside, we point out that all the terms that are proportional to
$i k_r$, as opposed to $k_r^2$, in equation
(\ref{eq:disp_full_nodim_kr}) are also proportional to some factor
$\epsilon_i$ with $i=1, 2, 3, 4$. All of these terms are negligible
for sufficiently small ratios $k_r/k_z$ no matter how strong the
toroidal field is.  Indeed,  if we consider perturbations with small
enough radial wavelengths, at some point, curvature effects will not
be important. However, this is not true in the vertical direction. In
that case, we can ignore the curvature terms only when the toroidal
field is weak.

\section*{\textsc{APPENDIX B}}
\label{sec:APPENDIX B}

In the various sections of the present study, we have seen that
the toroidal component of the magnetic field does play a role
in determining the stability criteria. In fact, for superthermal 
fields and quasi toroidal configurations. it dictates the values of
some of the critical wavenumbers for the onset of instabilities
(see \S \ref{subsec:analytic approximations}).
A particular, simple case, in which the importance of
considering both compressibility and magnetic tension terms can be
appreciated, is the study of modes with negligible frequency, i.e.,
with $\omega \ll 1$. We can obtain the wavenumber of these modes  by imposing
$\omega = 0$ to be a solution of the dispersion relation
(\ref{eq:disp_full_nodim_kr}). We obtain, in physical dimensions,
\begin{equation}
\label{eq:k_critical}
(k_z^0)^2 = - 2 \left.\frac{d \ln \Omega}{d \ln r}\right|_0  \left(
\frac{\Omega_0 r_0}{v_{{\rm A} z}} \right)^2  + 2 \epsilon_1
\epsilon_3  \left(\frac{v_{{\rm A} \phi}}{v_{{\rm A} z}} \right)^2 +
\epsilon_2 \epsilon_3 \left(\frac{v_{{\rm A} \phi}}{v_{{\rm A} z}}
\right)^2 \left(\frac{v_{{\rm A}\phi}}{c_{\rm s}} \right)^2 ~.
\end{equation}

In what follows, let us consider a rotationally supported disk whose
rotational profile is not too steep  (i.e., $|d \ln \Omega / d \ln r|$
is of order unity).  The \Alfven speed $v_{{\rm A} z}$ appears in all
three terms on the right hand side of equation (\ref{eq:k_critical})
and therefore it does not play a role in determining the relative
magnitudes between them.  Unlike the second term, the third term is
not necessarily small  with respect to the first one, when
superthermal fields are considered.  In this case, it seems again
(see \S \ref{subsec:analytic approximations}) safe
to neglect the curvature term proportional to  $\epsilon_1 \delta
v_{{\rm A}\phi}$ in equation  (\ref{eq:pert_euler_r_grads}). However,
had we neglected the curvature term proportional to $ \epsilon_2
\delta \rho$ in equation (\ref{eq:pert_euler_r_grads}) or the one
proportional to $\epsilon_3 \delta v_{{\rm A} r}$  in equation
(\ref{eq:pert_euler_phi_grads}), we would have missed the important
impact that the third term in equation (\ref{eq:k_critical}) has on
the stability of modes with $\omega \rightarrow 0$,  in the limit of
strong toroidal fields (see Fig. \ref{fig:unstable_regions_num}).
This is somewhat counterintuitive because there does not seem to be
any \emph{a priori} indication about which of the magnetic tension
terms (related to the curvature of the toroidal field component) is
less relevant in the original set of equations
(\ref{eq:pert_cont_grads})-(\ref{eq:pert_induc_z_grads}) for the
perturbations. This particular example illustrates the risks
associated with neglecting terms that are not strictly $2^{\rm nd}$
order in the perturbed quantities but rather address the 
geometrical characteristics of the 
background in which the (local) analysis is being carried out.

From equation (\ref{eq:k_critical}) it is also straightforward to see
under which circumstances it is safe to neglect the curvature terms.
For subthermal fields, $k_z^0$ will not differ significantly from
$k_{\rm BH}$ (eq.~[\ref{eq:k_BH}]) regardless of the geometry of the
field configuration. This is because, if the field is weak enough
($v_{{\rm A}} \ll c_{\rm s}$), no matter how strong of a (subthermal)
$B_\phi$ component we consider, the second and third term are
negligible with respect to the first one.  We conclude this short
analysis by commenting that, while a strong vertical field plays a
stabilizing role, in the sense that it drives $k_z^0$ toward small
values leaving all modes with shorter wavelengths stable, the
consequences of considering strong toroidal fields is a little more
subtle as it can be seen in the evolution of the structure of the
modes in Figure \ref{fig:comp_real}.

\acknowledgments{We thank Eliot Quataert, Jim Stone, Charles Gammie,
and Wolfgang Duschl for useful comments and discussion in different
stages of this study. We thank an anonymous referee for
pointing out the connection between instability II and the ATB modes.
We are also grateful to Ethan Vishniac for encouraging us to address 
the limit discussed in Appendix A. We also thank the Institute  for Advanced
Study for their hospitality during part of this  investigation.
This work was partially supported by NASA grant NAG-513374}

\end{document}